\documentstyle[psfig]{mn}

\newif\ifAMStwofonts

\ifoldfss
  \ifCUPmtlplainloaded \else
    \NewTextAlphabet{textbfit} {cmbxti10} {}
    \NewTextAlphabet{textbfss} {cmssbx10} {}
    \NewMathAlphabet{mathbfit} {cmbxti10} {} 
    \NewMathAlphabet{mathbfss} {cmssbx10} {} 
  \fi
  \ifAMStwofonts
    \ifCUPmtlplainloaded \else
      \NewSymbolFont{upmath} {eurm10}
      \NewSymbolFont{AMSa} {msam10}
      \NewMathSymbol{\upi}     {0}{upmath}{19}
      \NewMathSymbol{\umu}     {0}{upmath}{16}
      \NewMathSymbol{\upartial}{0}{upmath}{40}
      \NewMathSymbol{\leqslant}{3}{AMSa}{36}
      \NewMathSymbol{\geqslant}{3}{AMSa}{3E}

      \let\leq=\leqslant \let\le=\leqslant
      \let\geq=\geqslant \let\ge=\geqslant
    \fi
  \fi
\fi 

\ifnfssone
  \newmathalphabet{\mathit}
  \addtoversion{normal}{\mathit}{cmr}{m}{it}
  \addtoversion{bold}{\mathit}{cmr}{bx}{it}
  \newmathalphabet{\mathbfit} 
  \addtoversion{normal}{\mathbfit}{cmr}{bx}{it}
  \addtoversion{bold}{\mathbfit}{cmr}{bx}{it}
  \newmathalphabet{\mathbfss} 
  \addtoversion{normal}{\mathbfss}{cmss}{bx}{n}
  \addtoversion{bold}{\mathbfss}{cmss}{bx}{n}
  \ifAMStwofonts
    \ifCUPmtlplainloaded \else
      \UseAMStwoboldmath
      \makeatletter
      \new@mathgroup\upmath@group
      \define@mathgroup\mv@normal\upmath@group{eur}{m}{n}
      \define@mathgroup\mv@bold\upmath@group{eur}{b}{n}
      \edef\UPM{\hexnumber\upmath@group}
      \new@mathgroup\amsa@group
      \define@mathgroup\mv@normal\amsa@group{msa}{m}{n}
      \define@mathgroup\mv@bold\amsa@group{msa}{m}{n}
      \edef\AMSa{\hexnumber\amsa@group}
      \makeatother
      \mathchardef\upi="0\UPM19
      \mathchardef\umu="0\UPM16
      \mathchardef\upartial="0\UPM40
      \mathchardef\leqslant="3\AMSa36
      \mathchardef\geqslant="3\AMSa3E

      \let\leq=\leqslant \let\le=\leqslant
      \let\geq=\geqslant \let\ge=\geqslant
    \fi
  \fi
\fi 

\ifnfsstwo
  \DeclareMathAlphabet{\mathbfit}{OT1}{cmr}{bx}{it}
  \SetMathAlphabet\mathbfit{bold}{OT1}{cmr}{bx}{it}
  \DeclareMathAlphabet{\mathbfss}{OT1}{cmss}{bx}{n}
  \SetMathAlphabet\mathbfss{bold}{OT1}{cmss}{bx}{n}
  \ifAMStwofonts
    \ifCUPmtlplainloaded \else
      \DeclareSymbolFont{UPM}{U}{eur}{m}{n}
      \SetSymbolFont{UPM}{bold}{U}{eur}{b}{n}
      \DeclareSymbolFont{AMSa}{U}{msa}{m}{n}
      \DeclareMathSymbol{\upi}{0}{UPM}{"19}
      \DeclareMathSymbol{\umu}{0}{UPM}{"16}
      \DeclareMathSymbol{\upartial}{0}{UPM}{"40}
      \DeclareMathSymbol{\leqslant}{3}{AMSa}{"36}
      \DeclareMathSymbol{\geqslant}{3}{AMSa}{"3E}

      \let\leq=\leqslant \let\le=\leqslant
      \let\geq=\geqslant \let\ge=\geqslant
    \fi
  \fi
\fi 

\ifCUPmtlplainloaded \else
  \ifAMStwofonts \else 
    \def\upi{\pi}
    \def\umu{\mu}
    \def\upartial{\partial}
  \fi
\fi

\title[SN IIP 1999em in NGC 1637]{PHOTOMETRY AND SPECTROSCOPY OF THE \\TYPE IIP SN 1999em FROM OUTBURST TO \\DUST FORMATION. }

\author[Elmhamdi et al.]
       {Elmhamdi, Abouazza$^{1}$, Danziger, I.J.$^{2}$, Chugai, N.$^{6}$, Pastorello, A.$^{3,5,7}$ \and Turatto, M.$^{3}$, Cappellaro, E.$^{4}$, Altavilla, G.$^{3,5}$, Benetti, S.$^{3}$, Patat, F.$^{8}$, Salvo, M.$^{9}$\\
         $^{1}$SISSA / ISAS , via Beirut 4 - 34014 Trieste, Italy\\
	 $^{2}$Osservatorio Astronomico di Trieste, via G.B. Tiepolo 11,
         34131 Trieste, Italy\\
	 $^{3}$Osservatorio Astronomico di Padova, vicolo dell'Osservatorio 5,
         35122 Padova, Italy\\
         $^{4}$Osservatorio Astronomico di Capodimonte, via Moiariello 16,
         I-80131 Napoli, Italy\\
      	 $^{5}$Dipartimento di Astronomia, Universit\`a di Padova,
         vicolo dell'Osservatorio 2, 35122 Padova, Italy\\
	 $^{6}$Institute of Astronomy, Russian Academy of sciences, Pyatnitskaya 	48, 109017 Moscow, Russia\\
         $^{7}$Present address: Department of Physics and Astronomy-University 		of Oklahoma 440 W. Brooks St. ,Norman, OK 73019 (USA)\\
         $^{8}$European Southern Observatory, Karl Schwarzschild-Str. 2, D-8574		8, Garching bei M\"unchen, Germany\\
	 $^{9}$Mt.Stromlo Observatory, Cotter Road, 2611 Weston Creek, ACT, Australia}
\date{Accepted .....;
      Received ....;
      in original form ....}

\pagerange{\pageref{firstpage}--\pageref{lastpage}}
\pubyear{2002}

\begin{document}

\maketitle

\label{firstpage}

\begin{abstract}
We present photometry and spectra of the type IIP supernova 1999em 
in NGC 1637 from several days after the outburst till day
642. The radioactive tail of the recovered bolometric light curve of SN 1999em 
 indicates the amount 
of the ejected $^{56}$Ni to be $\approx 0.02~M_{\odot}$. The 
 H$\alpha$ and He I 10830~\AA\ lines at the nebular epoch show that the distribution of the bulk of $^{56}$Ni can be represented approximately 
by a sphere of $^{56}$Ni with a velocity 
of 1500 km s$^{-1}$, which is shifted towards the far hemisphere 
by about 400 km s$^{-1}$. The fine structure of the H$\alpha$ 
at the photospheric epoch reminiscent of the ``Bochum event" in SN~1987A 
 is analysed in terms of 
two plausible models: bi-polar $^{56}$Ni jets and 
non-monotonic behaviour of the H$\alpha$ optical depth combined with the 
one-sided $^{56}$Ni ejection. The late time 
spectra show a dramatic transformation of the 
[O I] 6300 \AA\  line profile between days 465 and 510, which we interpret 
as an effect of dust condensation during this period. Late time photometry supports the dust formation scenario after day 465. The [O I] line profile 
suggests that the dust occupies a sphere with 
velocity $\approx800$ km s$^{-1}$ and optical depth $\gg10$.
The latter exceeds the optical depth of 
the dusty zone in SN~1987A by more than 10 times. Use is made of the Expanding Photosphere Method to estimate the distance and the explosion time, $D \approx 7.83$ Mpc and  $t_{0}\simeq$ 1999 October 24.5 UT, in accord with observational constraints on the explosion time and with other results of detailed studies of the method (Hamuy et al. 2001; Leonard et al. 2002). The plateau brightness and duration combined 
with the expansion velocity suggest a presupernova radius of $120-150~R_{\odot}$,
 ejecta  mass of $10-11~M_{\odot}$ and explosion energy 
of $(0.5-1)\times10^{51}$ erg. The ejecta mass combined with the neutron star
and a conservative assumption about mass loss implies the main sequence 
progenitor of $M_{\rm ms}\approx 12-14~M_{\odot}$. The derived mass range is in agreement with the upper limit to the mass found using pre-supernova field images by Smartt et al. (2001). From the [O I] 6300,6364 \AA\ doublet 
luminosity we infer the oxygen mass to be 
a factor four lower than in SN~1987A which is consistent 
 with the estimated SN~1999em progenitor 
mass according to nucleosynthesis and stellar evolution theory. We note a ``second-plateau" behaviour of the light curve
 after the main plateau at the 
beginning of the radioactive tail. This feature seems to be common to SNe~IIP with 
low $^{56}$Ni mass.
\end{abstract}


\section{Introduction}
Type II (and Ib/Ic) SNe are generally associated with regions of recent star formation in spiral galaxies, suggesting that they represent the final episode in the life of massive stars ($M>8$$M_{\odot}$) which explode owing to core collapse \shortcite{fil2}.
The study of core collapse events is important for understanding the range of progenitor masses which produce them, the consequent nucleosynthesis for its effect on galactic chemical evolution and the explosion energy which remains an ill-determined quantity for the vast majority of SNe, and which is also relevant for gas dynamical processes and ejection of material from galaxies.
Detailed photometric and spectroscopic observations of SNe~II on a 
long time scale are still rare, especially for SNe~IIP (plateau).
Meanwhile from the recent experience with SN~1987A we know, how 
valuable can be extended sets of photometric and spectroscopic data for 
understanding what really happens 
to the massive ($M>8M_{\odot}$) star when its 
central iron core collapses. In this way one obtained information on 
the amount of ejected $^{56}$Ni and oxygen, and their mixing
and clumpiness, possible asymmetry of $^{56}$Ni ejecta, ``Bochum event", 
dust formation. 
The case of SN~1987A also provided us with the possibility for testing and 
revising the theory of stellar evolution, 
SN hydrodynamical models and models of spectra formation.

Supernova 1999em, discovered by the Lick observatory Supernova Search on Oct 29 UT at an unfiltered magnitude $\sim 13.5$ mag in the nearby galaxy NGC 1637 ($D=7.8$ Mpc; IAUC 7294), has become another well observed type IIP event. 
Detected very soon after the explosion and followed for more that 
600 days this event gives another boost to studies of SNe~II. SN 1999em is the first type IIP detected at both radio and X-ray wavelengths at early time \shortcite{pool}. However it is the least radio luminous and among the least X-ray luminous SNe ($\it Chandra$ X-ray and radio NRAO observations; IAUC 7318, 7336 ). This early and weak radio emission is consistent with a low mass loss shortly before the explosion for SNe IIP \shortcite{baron}. The analysis of the spectra at the early photospheric epoch 
already emphasized the problem of He abundance in the hydrogen envelope of 
SN~IIP progenitors (Baron et al. 2000) and permitted one to check and 
upgrade the method of the expanding photosphere (EPM) for the distance 
determination (Hamuy et al. 2001; Leonard et al. 2002). The spectropolarimetry on the other hand has been studied at 5 different epochs (till day 163 after discovery; Leonard et al. 2000). The authors estimated the broadband polarization at day 7 to be $\sim 0.2$ $\%$, rising to $\sim 0.5$ $\%$ on day 161. The low polarization found for SN 1999em and hence the small implied asphericity especially at early photospheric phase is another encouraging reason for the validity of the EPM for this class of object (SNe IIP). The progenitor nature of SN 1999em was discussed by Smartt et al. (2002), who used pre-explosion $CFHT$ images to derive bolometric luminosity limits and thus constraints on the mass of the progenitor star of SN 1999em. They concluded that the main-sequence mass should be $< 12 \pm1 M_{\odot}$.   

Here we present photometry and spectroscopy of SN~1999em from the 
phase of several days after the explosion till day 642. In what follows we describe the 
photometry and spectral evolution and provide estimates of the $^{56}$Ni and oxygen mass in SN~1999em; 
we will analyse fine structures of the line profiles 
at the photospheric and nebular epoch in an attempt to determine the 
effects of the $^{56}$Ni mixing and asymmetry. The last spectra are used as diagnostics of possible dust formation in the ejecta.
 The photometry and spectroscopy at the plateau phase 
will be used to recover global parameters of SN (presupernova radius, mass 
and energy of SN). We then discuss the main sequence mass of the 
progenitor and some important correlations.
\begin{figure}
\centerline{\psfig{file=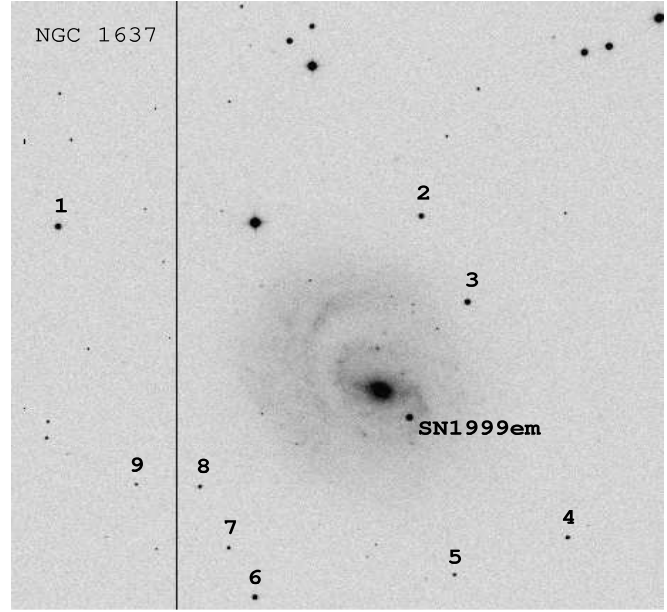,width=9truecm,height=9truecm}}
\caption{ SN 1999em in NGC 1637. Image taken at ESO on Mar 27. 2000, about 150 days after discovery (in the R band). The SN position is indicated, also shown are the reference sequence stars. North is up, East is to the left.}
\end{figure}
\section{Photometric Evolution and bolometric light curve}
Figure 1 shows the position of SN 1999em in its host galaxy NGC 1637. It is located at R.A=4:41:27.13, a Decl=-02:51:45.2, about $15''$ west and $17''$ south of the nucleus of its host galaxy NGC 1637. NGC 1637 is an SABc late type spiral galaxy having a heliocentric recession velocity of V$_{\rm hel}\sim$ 717 km s$^{-1}$. According to Sohn $\&$ Davidge \shortcite{sohn} it has a distance of  $7.8^{+1.0}_{-0.9}$ Mpc. A KAIT\footnote{0.8-m Katzman Automatic Imaging Telescope } image of the field taken on Oct 20.45 showed nothing at the position of SN 1999em, limiting the magnitude to about 19.0, suggesting that it was discovered shortly after the explosion.
\subsection{Interstellar extinction }
\begin{figure*}
\psfig{file=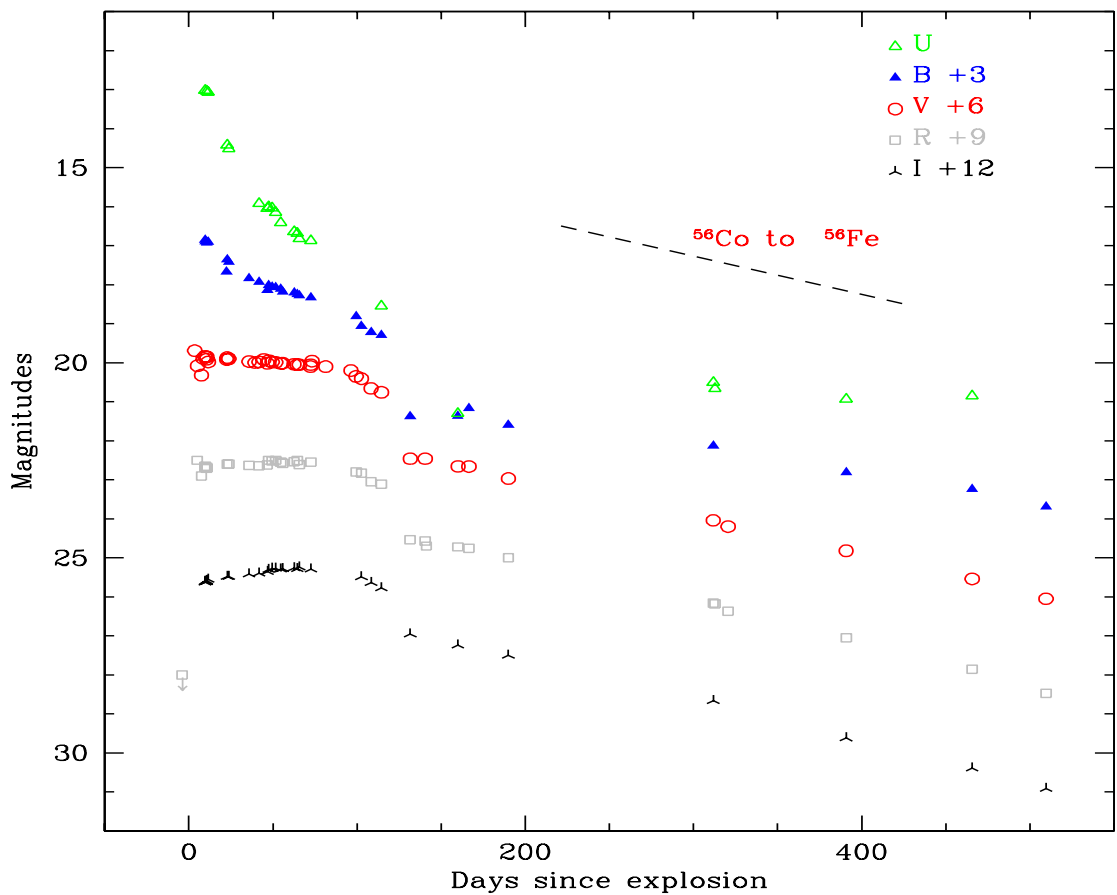,width=16cm,height=13cm}
\caption{ UBVRI Light curves of SN 1999em. The light curves have been shifted by the reported amounts. Also shown is the slope of $^{56}$Co to $^{56}$Fe decay. Reddening corrections have not been applied.}
\end{figure*}
Galactic extinction towards NGC 1637 is known to be $A_V^{\rm gal}=0.134$ \shortcite{schl} corresponding to a colour excess $\rm E(B-V)$=0.04 (adopting the standard reddening laws of Cardelli et al. 1989). However estimating the extinction due to the host galaxy is more complicated.

Recently Baron et al. (2000) have provided constraints on the total colour excess using theoretical modelling of two early time spectra, namely their 29 Oct $\&$ 4-5 Nov spectra. They found that $\rm E(B-V)=0.05-0.10$ and $\rm E(B-V)$ $\leq$ 0.15. A value of $\rm E(B-V)$=0.05, as derived from their SYNOW fit to their Oct 29 spectrum, implies an amount of $\rm E(B-V)$=0.01 due to the host galaxy corresponding to $A_V^{\rm host}=0.031$ and $A_B^{\rm host}=0.041$. Adopting, for example, these values one finds $M_{\rm max}^{\rm B} \leq -15.8$, which places SN 1999em at the lower extremity of the regular type II-plateau SNe in the classification of Patat et al. \shortcite{patat1}.

On the other hand we note the presence of a narrow Na I D interstellar absorption line in our first spectra at the redshift of NGC 1637 with an equivalent width (EW) $\sim$ 1.3 \AA, and therefore we would expect significant additional reddening in the host galaxy. In fact some correlations in the literature relating the equivalent width of Na I D to the reddening within the host galaxy indicate values of the order $\rm E(B-V)$ $\sim 0.32$ \shortcite{barb1} and $\rm E(B-V)$ $\sim 0.17$ (Benetti et al., unpublished correlation), corresponding to $A_V^{\rm host}=0.99$ and $A_V^{\rm host}=0.52$, respectively. If, for instance, we assume this last value, we obtain $M_{\rm max}^{\rm B} \leq -16.48$, which implies that SN 1999em is a regular SN IIP but with significant reddening. Uncertainties in the calibration of absorption line strengths prevent a confident determination of reddening by this method.

To overcome this disagreement between theoretical modelling and empirical calibration, Hamuy et al. \shortcite{ham3} have tested the sensitivity of the Expanding Photosphere Method (EPM) to reddening caused by the host galaxy in deriving the distance to the supernova (using different filter subsets, their Figure 12). They conclude a consistent value for the extinction in the host galaxy is $A_V^{\rm host}=0.18$ and thus a total extinction $A_V^{\rm tot}=0.31$ and $A_B^{\rm tot}=0.41$, which we adopt as the best estimate for our present analysis.
\subsection{Light curves }
\begin{table*}
\centering
\begin{minipage}{140mm}
  \caption{ SN 1999em Photometric Observations}
\begin{tabular}{c c c c c c c c}
\hline \hline
 Date &      JD   &    U($\sigma$)& B($\sigma$) & V($\sigma$) & R($\sigma$) & I($\sigma$)& Instrument\\
(UT) &2400000+&&&&&&\\
\hline        
28/10/99   &  51479.51 & ---     & ---     &    13.69 .06   & ---     & --- & 0\\    
03/11/99    &  51485.62 & 13.02 .03&  13.85 .03&   13.84 .03 &  13.67 .03&  13.63  .03 &4\\   
03/11/99    &  51485.81 &  ---    &  13.89 .03&   13.84 .03 &  13.65 .03&  13.61  .03 &2\\ 
04/11/99    &  51486.62 & 13.04 .03&  13.90 .03&   13.91 .03 &  13.71 .03&  13.59  .03 &4\\    
04/11/99    &  51486.73 &  ---    &  13.93 .03&   13.85 .03 &  13.68 .03&  13.61  .06 &2 \\ 
05/11/99    &  51487.57 & 13.07 .12&  13.90 .06&   13.98 .03 &  13.83 .09&  13.55  .12 &4\\   
05/12/99    &  51517.80 & 15.92   & 14.93  &  13.98  &  13.64  &  13.40 &1\\     
08/12/99    &  51520.56 &  ---     &  ---     &  13.92 .12 &  ---      &   --- &5\\
11/12/99   &  51523.54 & 16.01 .03 & 15.01 .03 &  13.95 .03 &  13.50 .03 & 13.30  .03&4 \\  
13/12/99   &  51525.51 & 16.04 .03 & 15.06 .03 &  13.98 .03 &  13.52 .03 & 13.27  .03&4 \\  
15/12/99   &  51527.57 & 16.15 .03 & 15.07 .03 &  13.99 .03 &  13.50 .03 & 13.27  .03&4 \\   
18/12/99   &  51530.73 & 16.41 .27 & 15.11 .09 &  14.02 .03 &  13.55 .03 & 13.29  .03&1 \\  
25/12/99   &  51538.59 & 16.64 .06 & 15.20 &  14.04 &  13.54 &  13.26  &4 \\  
27/12/99   &  51540.50 & 16.68 .03 & 15.25  &  14.05 &  13.51 &  13.28 &4\\   
28/12/99   &  51541.71 & 16.82 .09 & 15.28 .06 &  14.05 .03 &  13.61 .03&  13.24  .03&2\\   
05/01/00     &  51548.51 & 16.87 .03 & 15.33 .03 &  14.05 .03 &  13.55 .03&   13.29 .03&4 \\   
31/01/00   &  51575.44 &  ---     &  15.81 .12&   14.35 .09&   13.80 .03&  ---&5 \\   
09/02/00   &  51584.32 &  ---     & 16.22 .39 &  14.66 .09 &  14.05 .06 & 13.64  .06 &5\\  
16/02/00    &  51590.53 & 18.55 & 16.29 &  14.76 &  14.11 & 13.77  &2 \\  
13/03/00    & 51616.50  & ---      & ---      & 16.46 .06  & 15.57 .06 & ---  &  2 \\ 
08/04/00    &  51642.52 &  ---     & 18.17 &  16.66  &  15.76 &  ---&1\\ 
31/08/00    &  51787.71 &  ---     &  ---     &  18.04 .06 &  17.16 .03 &  ---  &3 \\
31/08/00    &  51787.74 & 20.51 .21 & 19.13 .06 &   ---     &   ---     &   16.66  .06&3\\       
01/09/00    &  51788.90 & 20.66 1.05 &  ---     &   ---     &  17.18 .06 & ---  & 1\\
08/09/00     &  51796.60 &  ---     &  ---     &  18.20 .15 &  17.37 .09 & --- & 5\\ 
17/11/00   &  51866.77 & 20.94 .72 & 19.81 .54 &  18.82 .24 &  18.05 .12 & 17.62  .24&2\\   
01/02/01   &  51941.65 & 20.85 .12 & 20.24 .18 &  19.54 .12 &  18.86 .12 & 18.39  .09&1 \\   
16/03/01   &  51985.53 &  ---     & 20.69 .27 &  20.05 .18 &  19.47 .24 & 18.91  .21&2 \\
  \hline \hline\\
\end{tabular} 
\end{minipage}
\footnotesize \\
\emph{\rm 0 = WHT(Image provided by Smartt.); 1 = ESO 3.6m+EFOSC2; 2 = Danish 1.54m+DFOSC \\ 3 = TNG+DOLORES; 4 = TNG+OIG; 5 = Asiago 1.82m+AFOSC \\ \hspace*{-8.5cm}$\star$ The reported errors are 1$\sigma$.}
\normalsize
\end{table*}
The UBVRI photometry data are reported in Table 1 together with the different telescopes and instruments with which the observations were obtained. Our photometry started at $\sim$9 d and extended to $\sim$508 d (after explosion time). The reduction was carried out within the IRAF environment, applying bias, overscan and flat-field corrections. The contamination of the galaxy has been removed using PSF substraction. The photometric calibration of the SN was made relative to the local sequence of stars (see Figure 1) in the field of the host galaxy NGC 1637 and calibrated using observations of Landolt standard stars \shortcite{land}, obtained when the nights were photometric. The SN magnitudes have been measured using a PSF-fitting technique.

The UBVRI light curves of SN 1999em are illustrated in Figure 2. The Figure includes also data from Hamuy et al. (2001) as well some data from circulars and VSNET service\footnote{http://www.kusastro.kyoto-u.ac.jp/vsnet/}. We have also plotted the decay slope of $^{56}$Co : 0.98 mag (100 d)$^{-1}$, which corresponds to the late-phase decline rate of most normal SNe II L-P, especially for the V-band (Turatto et al. 1990 and Patat et al. 1994). 

At early times, especially in the V band, one notes that SN 1999em rises to a brief peak, then drops by about 0.6 mag in $\sim3.5$ days, then rises again to a brief second maximum followed by a settling onto the plateau phase. Here the expansion of the photosphere and the cooling balance each other to produce a phase of almost constant magnitude, $m_{V}\sim 14$ and $M_{V}\sim -15.76$. The plateau phase lasts about 80 days, after which the light curve displays a steep decline ($\sim 2$ mag in $\sim 40$ days for the V-band), signaling the onset of the nebular phase and the start of the exponential decline. Similar early short duration rapid change was seen in SN IIP 1988A \shortcite{ruiz} as well as for SN 1993J \shortcite{barb2}, and can be interpreted to be a consequence of low mass loss in the immediate presupernova phase (see discussion of the bolometric light curve).

After the steep decline from the plateau phase and starting around 130 days after explosion the shapes of the light curves evolve with time: firstly there is a clear flattening from day 130 lasting about 50 days, especially at blue wavelengths, which is then followed by the exponential decline. Similar flattening is also seen in the data for the faint SN IIP 1997D \shortcite{ben} and is present in the light curves of SN IIP 1991G (Blanton et al. 1995). 

A linear fit to the tail (from $\sim$180 to $\sim$ 510) gives the following decline rates (in mag $(100\rm d)^{-1}$ ): $\gamma_{B}\sim 0.66$, $\gamma_{V}\sim 0.97$,  $\gamma_{R}\sim 1.08$ and $\gamma_{I}\sim 1.07$ which are close to the values found for the typical SN IIP 1969L (Turatto et al. 1990). Especially for the V-light curve this indicates that radioactive decay of $^{56}$Co with consequent trapping of $\gamma$-rays is the main source of energy powering the light curve at late times, at least until $\sim 510$ days after explosion.

We note also that the late U light curve remains almost flat, similar in behaviour to that reported for SN 1987A until $\sim$ 400 days after explosion \shortcite{sunt}.
\subsection{Colour evolution }
\begin{figure}
\psfig{file=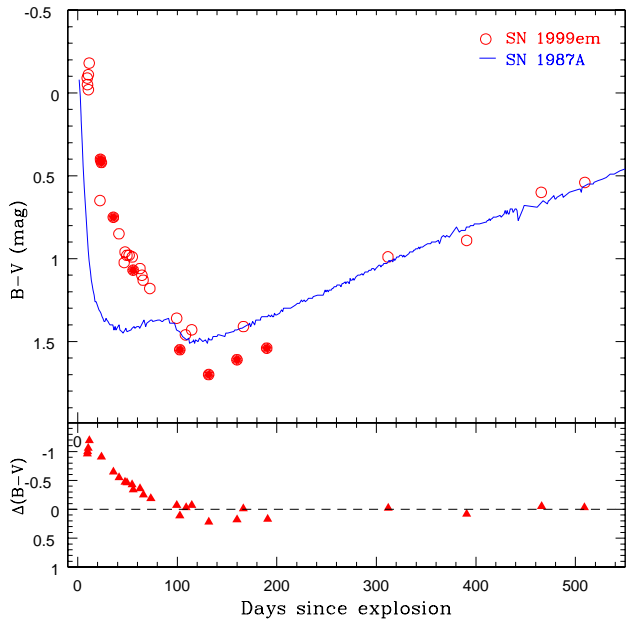,width=9cm,height=9cm}
\caption{$\rm B-V$ colour of SN~1999em and SN~1987A. Bottom panel shows the 
difference $\Delta(\mbox{B}-\mbox{V})=(\mbox{B}-\mbox{V})_{\rm 99em}-
(\mbox{B}-\mbox{V})_{\rm 87A}$. Both SNe have been corrected for reddening. For SN 1987A we use $A_{\rm V}=0.6$ \shortcite{fil1}. Filled circles represent data from Hamuy et al. 2001. }
\end{figure} 
In Figure 3 we show the $\rm (B-V)$ colour curve of SN 1999em, together with that of SN 1987A for comparison. The explosion date
of SN 1987A is well established to be February 23.316 from the Kamiokande and
IMB neutrino detections (Hirata et al. 1987; Bionta et al. 1987), while for SN 1999em the explosion date is assumed to be October 24.5 ($\sim$ JD 2451476) derived from the EPM, to be discussed later.

The $\rm (B-V)$ colour varies at different rates according to phase. During the first 40 days, it exhibits a steep and rapid decline from blue (high temperatures) to red (low temperatures) as the supernova envelope expands and cools. Subsequently the $\rm (B-V)$ colour varies more slowly as the rate of cooling decreases. At $\sim$130 d the $\rm (B-V)$ colour reaches a value around 1.7, then it turns blue again (rate $\sim -0.33~\rm  mag~(100~d)^{-1}$), as the light curve settles into the exponential tail.

Note the general similarity of $\rm (B-V)$ colour evolution of SN 1999em with that of SN 1987A especially at late phases, where the $\Delta\rm (B-V)$ approaches zero (the bottom box in Figure 3). This fact gives us confidence that comparison of absolute V-light curves should be reliable for the estimation of $^{56}$Ni mass, released in the explosion of SN 1999em, because the bolometric light-curves are liable to be similar also.

There is little doubt that the difference of the colour behaviour of both 
supernovae at the 
early epoch (faster cooling of SN~1987A) is related to the difference 
in presupernovae radii. The smaller radius of SN~1987A presupernova leads to 
a faster adiabatic cooling of the radiation trapped in the expanding
envelope. The details of the colour behaviour of SN~1999em 
should be reproduced in future hydrodynamical modelling.
\begin{figure}
\psfig{file=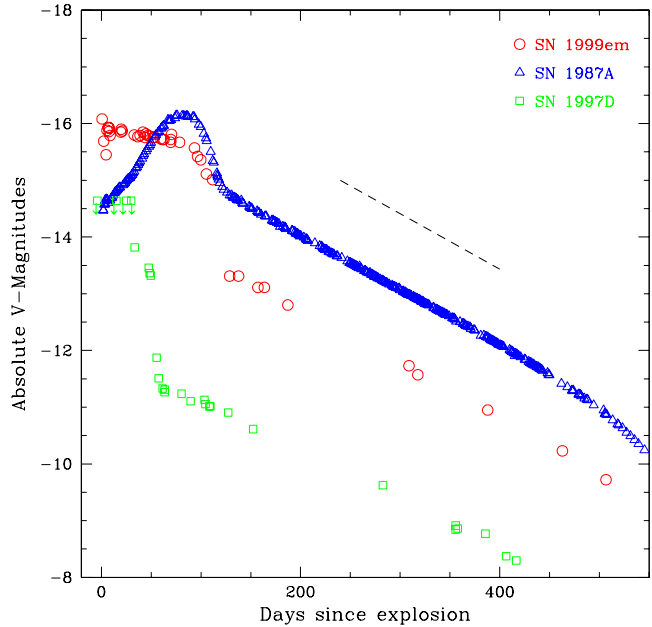,width=9truecm,height=9truecm}
\caption{ Comparison of the V absolute light curve of SN 1999em with those of SN 1987A and SN 1997D. Distances moduli of $\mu=29.46$, $\mu=18.49$ and $\mu=30.64$ are assumed for the three SNe, respectively. Dashed line shows the luminosity dependence due to $^{56}$Co decay for arbitrary mass of $^{56}$Ni.}
\end{figure}
\subsection{The bolometric light curve and $^{56}\rm Ni$ mass }
\begin{figure*}
\centerline{\psfig{file=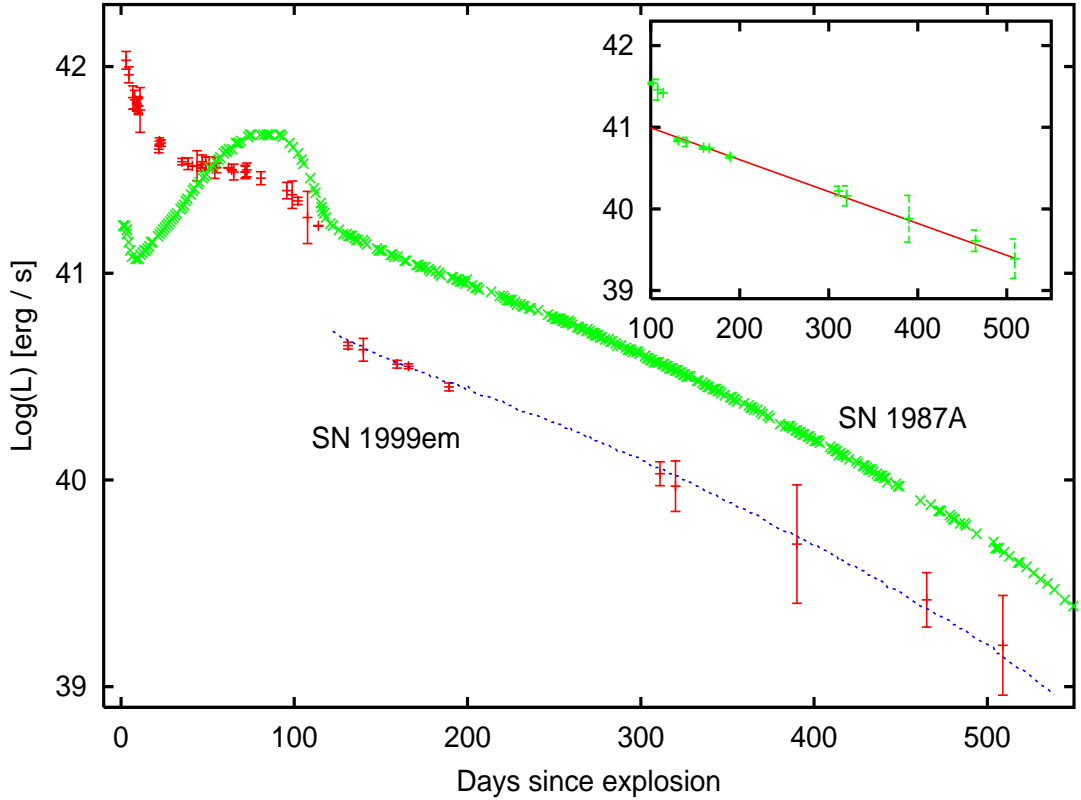,width=15truecm,height=11truecm}}
\caption{The UBVRI bolometric light curve of SN 1999em together with that of SN 1987A for comparison. The dotted line corresponds to the best fit of SN 1999em late data by SN 1987A tail. The window on the right shows the resulting tail scaled up by an amount of 0.19 dex (provided by integrated J,H,K fluxes) along with the $^{56}$Co decay for 0.02 of $^{56}$Ni. The reported error bars are 1$\sigma$.}
\end{figure*}
In Figure 4, we present the absolute light curve of SN 1999em in the V band. The light curve of the double-peaked SN 1987A and the peculiar type IIP SN 1997D are also displayed for comparison. These plots highlight the different behaviour of the photometric evolution of the three SNe, from the explosion until about 550 days. They provide constraints on the explosion energy of SN 1999em as well as on the radioactive $^{56}$Ni mass ejected.
All the radioactive tails of the three SNe, especially in the V band, follow the $^{56}$Co decay slope suggesting that it is the main energy source powering the late phases of the light curves.
The fainter object SN 1997D, $M_{\rm V}^{\rm max} \geq-14.65$ \shortcite{tura3}, ejected an extremely small amount of radioactive $^{56}$Ni, about 0.002 $M_{\odot}$, with a low explosion energy of $\sim 10^{50}$ ergs derived from modelling the spectra \shortcite{chug3}, while the mass of $^{56}$Ni ejected by the well studied SN 1987A and the explosion energy are $\sim$0.075 $M_{\odot}$ and $\sim 1.3\times10^{51}$ ergs, respectively (Danziger et al. 1988; Woosley et al. 1989). We therefore conclude that SN 1999em produced a mass of $^{56}$Ni intermediate between 0.002 and 0.075 $M_{\odot}$. To obtain a more accurate estimate of the $^{56}$Ni mass we constructed the ``bolometric'' light curve of SN 1999em by integrating the flux in the UBVRI bands. The resulting light curve is shown in Figure 5 together with that of SN 1987A. 

A least squares fit of the tail (from +140 d to +465 d) yields an $e$-folding time of $\sim119$ days. Comparing the bolometric luminosities of the tail of SN 1999em with SN 1987A we obtain a best fit of SN 1999em data to the tail of SN 1987A for $L(\rm{99em})/\it L(\rm 87A)=0.30$. With the $^{56}$Ni mass 
$0.075~M_{\odot}$ in SN~1987A we thus derive for SN~1999em the 
amount of ejected  $^{56}$Ni $\sim 0.022~M_{\odot}$. The bolometric luminosity should be corrected for the 
contribution of the flux in JHK bands. According to our photometric 
observation on day 440 the contribution of the JHKs flux is about
0.19 dex. Then assuming a similar constant percentage contribution by the infrared flux during the tail phase (which is not precise), we scale up the derived UBVRI light curve by 0.19 dex. The resulting late curve is shown in the window in Figure 5 together with the best theoretical fit of the radioactive decay energy input assuming that the envelope is optically thick to $\gamma$-rays ($L=L_{0}\times$($M_{\rm Ni}$/$M_{\odot}$)e$^{-t/\tau_{\rm Co}}$, with $L_{0}$=1.32$\times$ 10$^{43}$ ergs s$^{-1}$ and $\tau_{\rm Co}=111.23$ days; Woosley et al. 1989). The best fit is found for $^{56}$Ni mass $0.02\pm 0.0013~M_{\odot}$. Both methods thus lead to the consistent value of $^{56}$Ni mass of $0.02~M_{\odot}$. This amount is similar to the one derived for SN 1991G ($\sim$ 0.024 $M_{\odot}$; Blanton et al. 1995) and smaller than the mass determined for the typical SNe IIP 1969L and 1988A ($\sim$ 0.07 $M_{\odot}$; Turatto et al. 1993).
\section{Spectroscopic Evolution }
The journal of the spectroscopic observations in the optical as well as in the infrared are displayed in Table 2. For each spectrum the date (col.1), phase (col.2), range (col.3) and the instrument used (col.4) are presented. The spectra have been wavelength calibrated using comparison arc-spectra of He-Ar or He-Ne lamps, and flux calibrated using observations of spectrophotometric standard stars selected from Stone $\&$ Baldwin 1983, Baldwin $\&$ Stone 1984; Hamuy et al. 1992 and Hamuy et al. 1994.  
\begin{figure*}
\centerline{\psfig{file=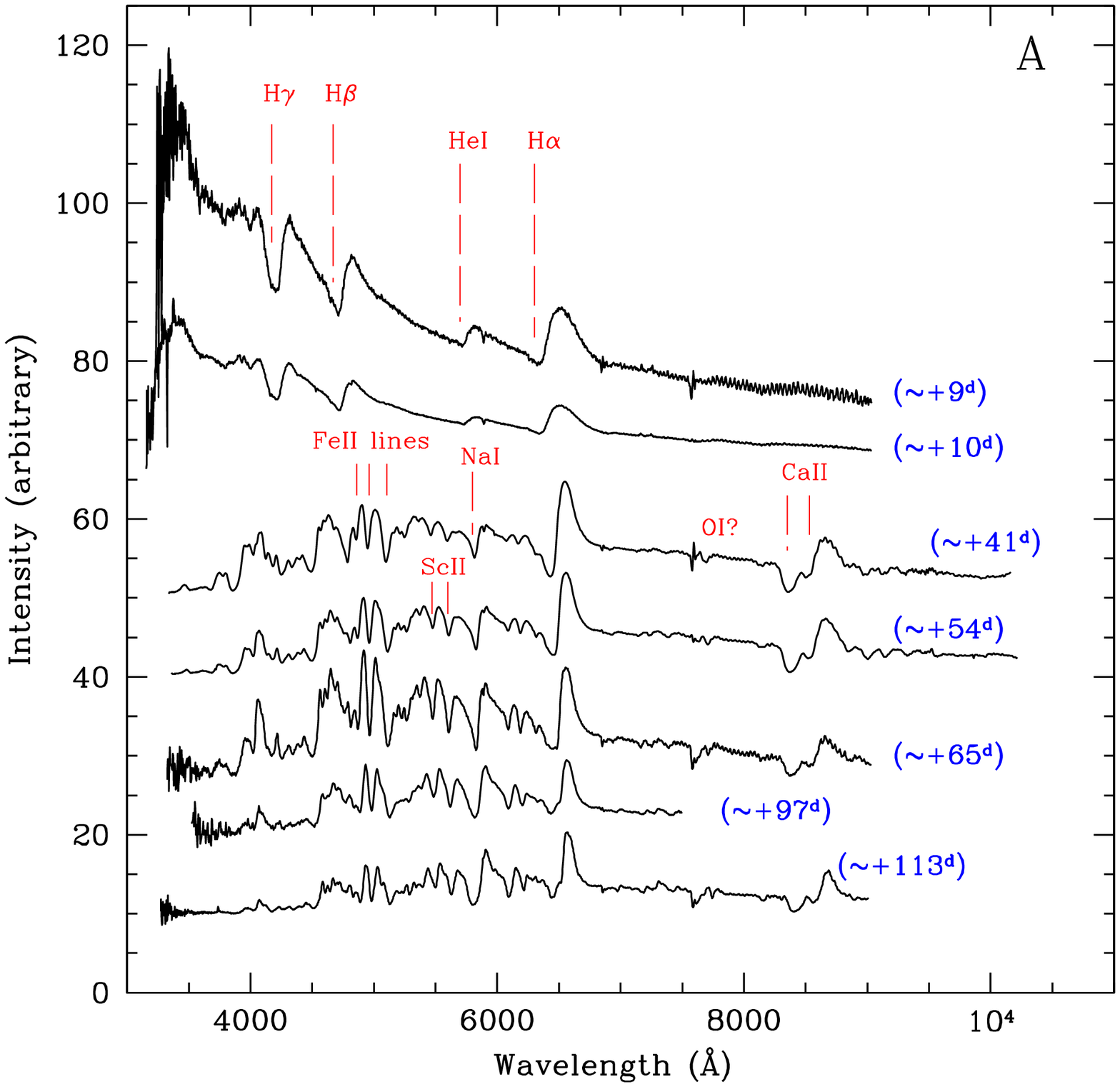,width=9cm,height=11cm}\psfig{file=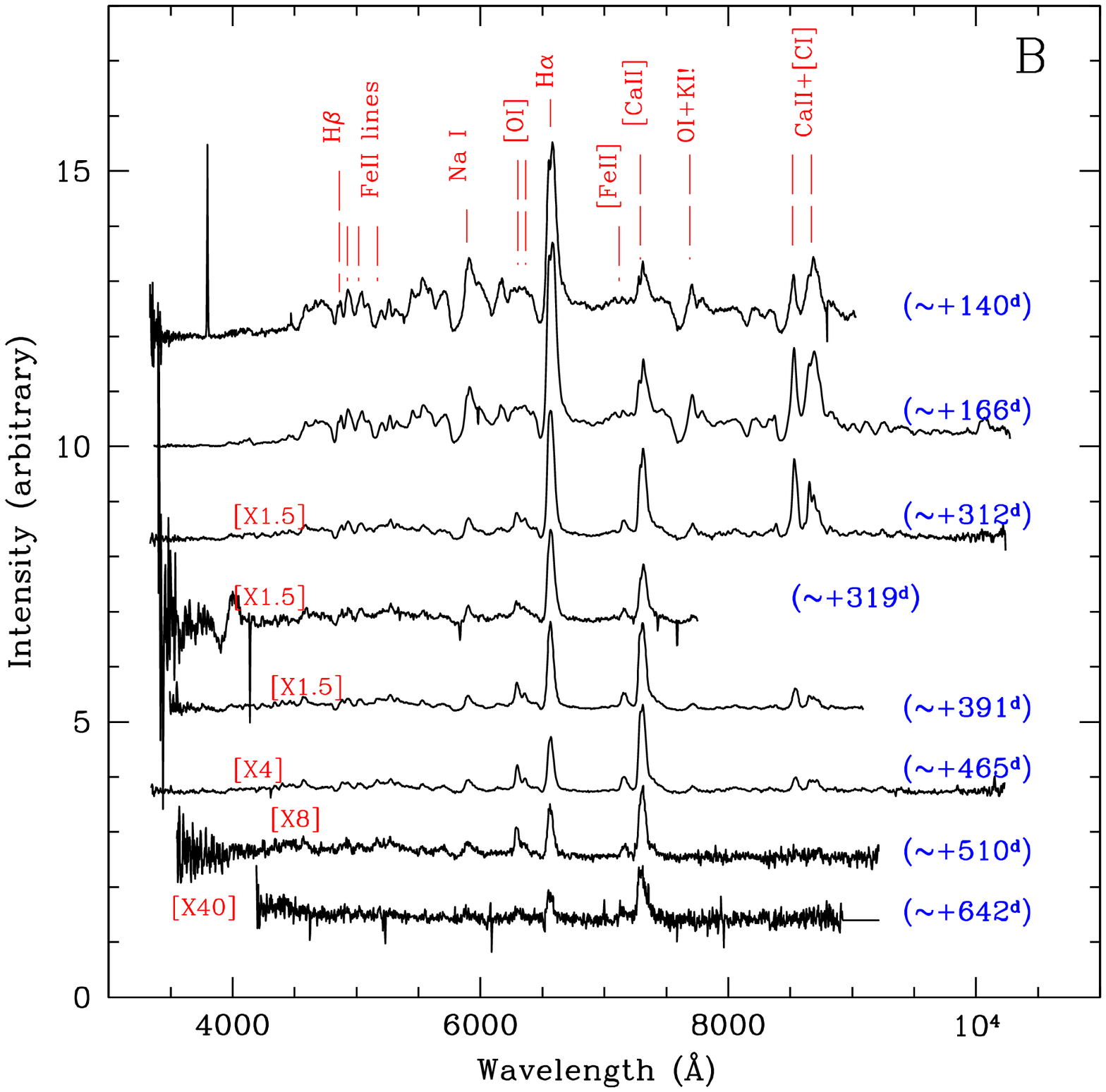,width=9cm,height=11cm}}
\caption{ Spectral evolution of SN 1999em during the photospheric phase(A), and during the nebular phase(B). All the spectra have been corrected for the recession velocity of the host galaxy NGC 1637 and vertically displaced by an arbitrary constants for clarity. Each spectrum is labelled by the corresponding days after explosion. The strongest features are also marked as absorption in the photospheric phase(A) and at the rest frame in the nebular phase(B).}
\end{figure*}
\begin{table} \caption{Spectroscopic observations of SN 1999em}
\begin{tabular}{c c c c }
\hline \hline
 Date (UT; dd/mm/yy) & Phase$^{\dagger}$   &  Range (\AA)& Instrument\\
\hline 
$\star$ Optical observations :\\
03/11/99   & $+$9 & 3240-9060  & DI \\
04/11/99  &$+$10   & 3160-9050& DI\\
05/12/99  &$+$41   & 3340-10190& EI\\ 
18/12/99  &$+$54  & 3370-10200& EI\\
29/12/99  &$+$65  & 3340-9050& DI\\
31/01/00  &$+$97  & 3530-7520& AI\\
16/02/00  &$+$113  & 3280-9030& DI\\
13/03/00  &$+$140   & 3340-9050& DI\\
08/04/00  &$+$166   & 3380-10300& EI\\
01/09/00  &$+$312   & 3350-10260& EI\\
08/09/00  &$+$319   & 3410-7770& AI\\
19/11/00  &$+$391   & 3500-9110& DI\\
01/02/01  &$+$465   & 3360-10260& EI\\
18/03/01  &$+$510   & 3560-9240& DI\\
27/07/01  &$+$642   & 4200-9940& EII\\
\hline
\hspace{-0.5truecm}$\star$ IR observations :\\
14/11/99  &$+$20   & 9420-25040& EIII\\
11/01/01  &$+$444  & 9430-16540& EIII\\
\hline \hline
\end{tabular} \\[1ex]
\footnotesize
\emph{\rm $\dagger$ since the explosion time: $\sim$ 24.5/10/1999\\ AI = Asiago 1.82m+AFOSC; DI = Danish 1.54m+DFOSC;\\ EI = ESO 3.6m+EFOSC2; EII = ESO-VLT-U1+FORS1 ;\\ EIII = ESO-NTT+SOFI  }\\
\normalsize
\end{table}
\subsection{Optical spectra}
\begin{figure*}
\psfig{file=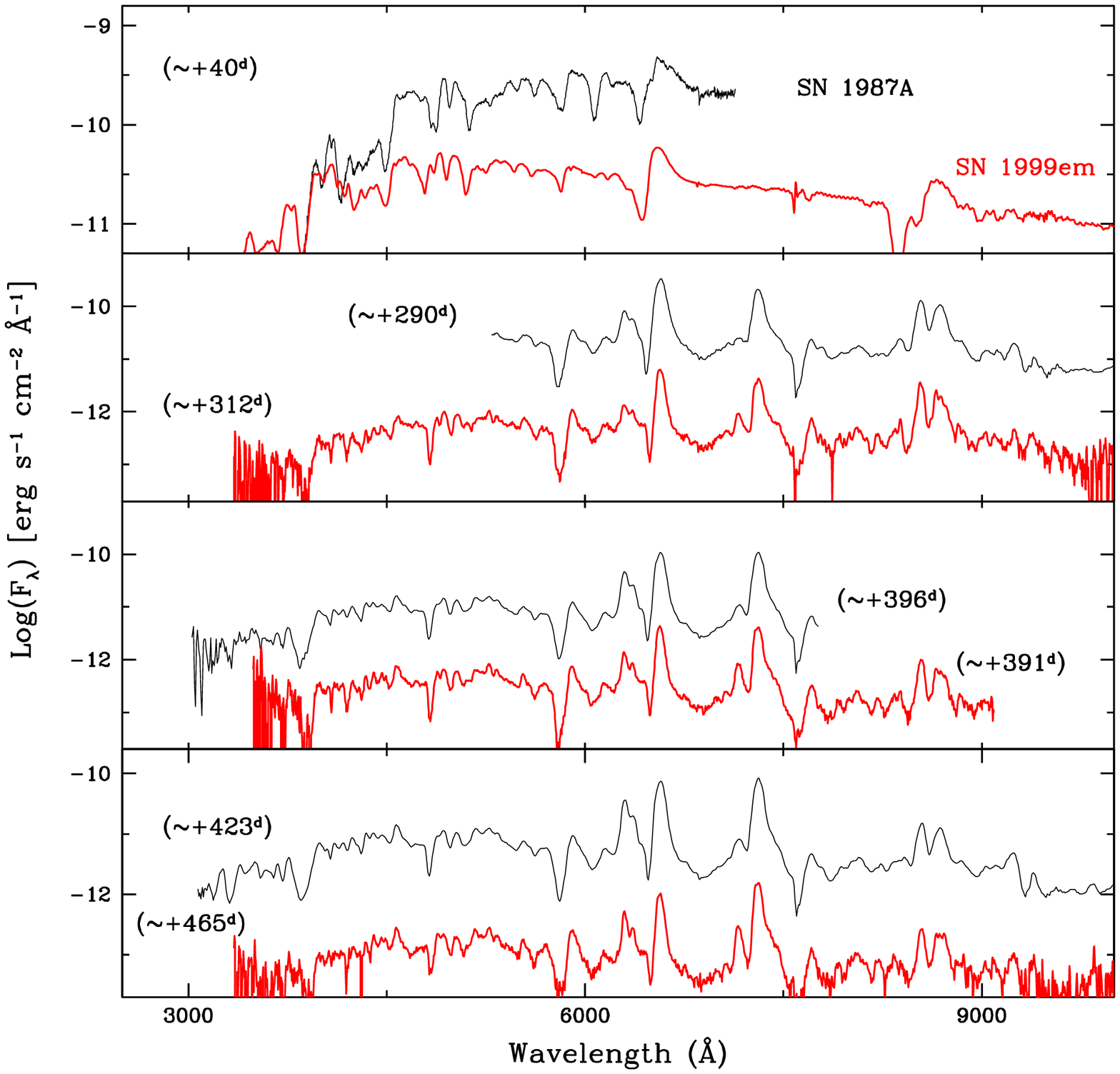,width=16truecm,height=15truecm}
\caption{ The spectral evolution of SN 1999em with that of SN 1987A at similar phases. The spectra of SN 1999em have been shifted by 3.5 dex (log-scale). The flux scales correspond to SN 1987A. The spectra are corrected for redshift as well for reddening by the same amounts as Fig. 3.}
\end{figure*}
We present the spectral evolution of SN 1999em which extends from $\sim+9~\rm d $ to $\sim+642~\rm d $ since explosion. The redshift $cz$=717 km s$^{-1}$ of the parent galaxy NGC 1637 has been removed in all spectra . 

In Figure 6A we show the evolution during the photospheric phase. The earliest spectra display a blue continuum indicating temperatures exceeding 10$^{4}$ K with broad P-Cygni profiles of the hydrogen Balmer lines, He I lines, and the Na I D and Ca II lines. In the first spectrum the blue wing of H$\alpha$ absorption indicate expansion velocities up to $\sim 16000$ km s$^{-1}$.

Those facts, broad P-Cygni profiles of hydrogen lines, high expansion velocities and high temperatures, imply that the supernova is a type II SN with a substantial hydrogen envelope discovered close to the explosion time. Note that the He I 5876 \AA\ with a P-Cygni profile is visible for the 3 and 4 Nov spectra (the same behaviour was seen in SN 1987A). In fact, in early time spectra Baron et al. (2000) have found evidence for helium enhanced by at least a factor of 2 over the solar value as well as a possible nitrogen enhancement. We do not find clear evidence for nitrogen lines in our first spectra. As to the He I line, the interpretation of its strength in terms of He overabundance is premature, since ignored freeze-out effects could 
lead to the enhanced He excitation compared to the steady-state model
(Utrobin \& Chugai 2002). 

At the middle of the plateau phase the photosphere
 cools down to $5000-6000$~K; at this 
epoch the spectra (Fig. 6A) are dominated 
by strong lines of H, Ca II, Fe II, Na I with well developed P-Cygni 
profiles which become narrower as the velocity at the photosphere 
decreases. We note also the appearance of absorption lines at 5476 \AA\ and 5608 \AA\ which we attribute to lines of Sc II 5526 \AA\ and 5658 \AA, respectively although secondary evidence from other lines of these ions is not obvious. 
As the supernova ages, the internal energy in the form of the trapped radiation is exhausted, the luminosity 
drops signaling the end of the photospheric epoch and transition to the 
nebular phase. The continuum formed in the region excited by $^{56}$Co decay 
becomes faint and diluted compared to black-body intensity. As a result the contrast between the net line emission and continuum
increases, which is demonstrated {\em e.g.} by H$\alpha$ and Ca II 8600 \AA\ triplet (last spectrum in Fig. 6A).

Figure 6B shows the spectra during the nebular phase, from $\sim+140~\rm d$ until $\sim+642~\rm d$ after explosion. The supernova is now in the radioactive tail phase. The first two spectra signal the onset of the nebular phase, where relatively broad P-Cygni profiles of H$\alpha$ and Na I 5890,5896 \AA\ were still seen with FWHM of H$\alpha$ about $ 3500$ km s$^{-1}$. About five months later, $\sim+312~\rm d$ after explosion, H$\alpha$ has become still narrower with FWHM of $\sim2400$ km s$^{-1}$. The feature at 7300 \AA\ is identified with 
[Ca II] 7392, 7324 \AA\ doublet always observed in nebular spectra 
of SNe~IIP. What is unusual is the notable domination of the [Ca II] 7300 \AA\ 
over H$\alpha$ in the latest spectra. Note also the emergence of the nebular emission of the [O I] 6300,6364 \AA\ doublet. Its profile evolution can be seen in Figure 6B. The $140~\rm d$ and $166~\rm d$ spectra showed already the presence of the [O I] although the two components were not yet resolved.
As time progresses the [O I] 6300 \AA\ component increases relative to the [O I] 6364 \AA\ one, indicating that the lines become optically thin as the supernova envelope expands.

The spectra of SN~1999em at photospheric and nebular epochs 
in many respects are similar to those of SN~1987A (Fig. 7). 
Note, however, the line width and hence velocities in SN 1987A are higher e.g. at $\sim+40~\rm d$: $V_{\rm 87A}(H\alpha)\sim 6900$ km s$^{-1}$, $V_{\rm 99em}(H\alpha)\sim 6100$ km s$^{-1}$ from the P-Cygni absorption minima. The Ba II 6142 \AA\ absorption line of an s-process element, is very prominent for SN 1987A, while its presence is only hinted in the spectrum of SN 1999em. The Sc II 6280 \AA\ line, if correctly identified, on the other hand has almost the same strength for the two spectra.
The 4554 \AA\ feature, partly due to Ba II 4554 \AA, is similar for the two SNe. This is possibly due to a blend with other lines, in particular Fe II 4555 \AA\ and Ti II 4549 \AA. 
Note also the emergence of the [Fe II] 7155 \AA\ emission line. Indeed we measure the maximum intensities of this line and comparing it with the strong [Ca II] 7291,7324 \AA\ line we obtain at $\sim390 ~\rm d$: $I_{\rm max}$([Ca II])/$I_{\rm max}$([Fe II]) = 6.98 for SN 1999em, while it is around $24.68$ for SN 1987A. This suggests that the [Fe II] lines do not originate from newly synthesized iron. 

In Figure 8 we report the luminosity evolution of some nebular lines, namely H$\alpha$, [Ca II] 7291,7324 \AA\ (upper panel) and [O I] 6300,64 \AA\ (lower panel) for SN 1999em. The same data for SN 1987A, at similar times, are also shown for comparison. The evolution of H$\alpha$ and [Ca II] 7291,7324 \AA\ are clearly decreasing in time for both supernovae in a similar and parallel way. This reflects the exponential decline at late times. On the other hand the derived luminosities for SN 1999em are lower than for SN 1987A which must be related to the nature and diversities in the progenitor stars of the two events. It is noteworthy that the ratio of  H$\alpha$ luminosity of SN 1999em and SN 1987A between days 300 and 400 is $\approx 0.3$, i.e. the same as the ratio of the $^{56}$Ni. This is perhaps not surprising since the rate of H$\alpha$ emission is determined by the rate of radioactive decay. However, we know that hydrogen photoionization from the second level is involved also in producing H$\alpha$ quanta (Chugai 1987; Xu et al. 1992), which, although it also is eventually determined by radioactive decay, might spoil the scaling of H$\alpha$ luminosity with $^{56}$Ni mass especially at the early nebular epoch. Moreover, at the early epoch collisional de-excitation probably plays a role also. On the other hand at later epoch ($>400$ d) the escape of $\gamma$-rays should be significant, with the escape rate dependent on the mass and explosion energy. Therefore the epoch $300-400$ days is just the right phase at which the H$\alpha$ luminosity scales as $^{56}$Ni mass in SNe IIP. This suggests to us the idea of using spectrophotometry of SNe IIP around day 300 in the H$\alpha$ band with low resolution (say 1000 km s$^{-1}$) to measure $^{56}$Ni mass based upon comparison of H$\alpha$ luminosity with the template SN 1987A, without the need to reconstruct the bolometric light curve, which would be difficult for distant SNe IIP.
\begin{figure}
\psfig{file=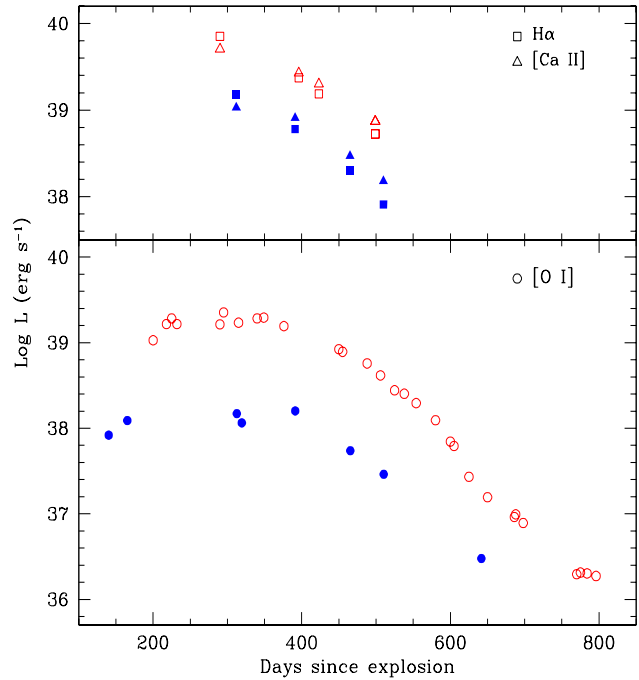,width=9truecm,height=10truecm}
\caption{ The temporal evolution of the luminosity of some nebular lines for SN 1999em with that of SN 1987A at similar times. Upper panel: the [Ca II]7291,7324 \AA\ and H$\alpha$ evolution. Lower panel: the [O I] 6300,64 \AA\ doublet evolution. Filled symbols correspond to SN 1999em while open symbols represent SN 1987A data.}
\end{figure}
\begin{figure}
\psfig{file=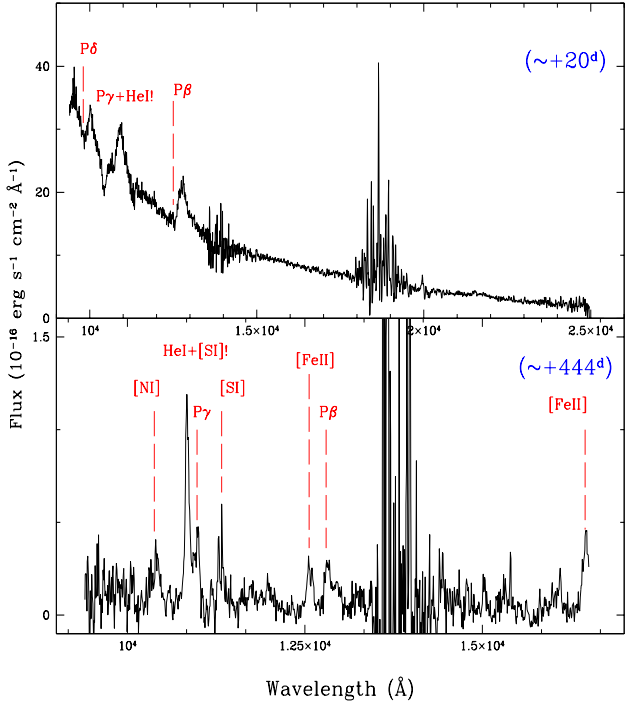,width=9cm,height=10cm}
\caption{ The infrared spectra on day 20 and 444 after correcting them from the redshift of the host galaxy. Some line identifications for both phases are also reported.}
\end{figure}

The behaviour of the luminosity of [O I] doublet in both supernovae 
is similar with a ``plateau-like'' maximum from $\sim166~\rm d$ until $\sim391~\rm d$ for SN 1999em followed by a clear steep decline later on, as for SN 1987A. We measured this decline rate for SN 1999em to be $\sim0.0166$ mag day$^{-1}$.

Yet the luminosity of [O I] doublet in SN~1999em is a factor of 15 lower 
between day $\sim300$ and $\sim500$ which indicates a lower amount
of oxygen compared to that of SN 1987A.
 To get a rough idea 
of the ratio of oxygen mass in both supernovae one
notes that the luminosity of [O I] doublet at the epoch of 
 $\sim 1$ year is powered by the $\gamma$-ray deposition and by
  ultraviolet emission arising from the deposition 
  of $\gamma$-rays in oxygen-poor material. Generally, one may 
 write the [O I] doublet luminosity as: 
\begin{equation}
\hspace{2cm} L(6300)=\eta \frac{M_{\rm O}}{M_{\rm ex}}L(^{56}\mbox{Co}),
\end{equation}  
where $M_{\rm O}$ is the mass of oxygen,
$M_{\rm ex}$ is the $``$excited" mass in which the bulk of 
radioactive energy is deposited, and $\eta$ is the efficiency 
of transformation of the energy deposited in oxygen into 
the [O I] doublet radiation. 
Given 15 times lower luminosity of [O I] doublet in 
SN~1999em (before dust formation) and factor 3.4 lower $^{56}$Ni mass in SN~1999em (section 2.4),
and assuming that in both supernovae $\eta$ and $M_{\rm ex}$ are 
similar, we derive the rough estimate
that the mass of oxygen in SN~1999em is a factor 4.4 lower than 
in SN~1987A. The oxygen mass in SN~1987A according to different 
determinations (Fransson \& al. 1993; Chugai 1994) is in the range 
 $1.5-2$ $M_{\odot}$, which translates into $\sim 0.3-0.4$ $M_{\odot}$ 
of oxygen in SN~1999em. Given the nucleosynthesis
computations (Woosley \&  Weaver 1995) this mass corresponds 
to the main-sequence stellar mass of $13-14~M_{\odot}$.
\begin{figure}
\psfig{file=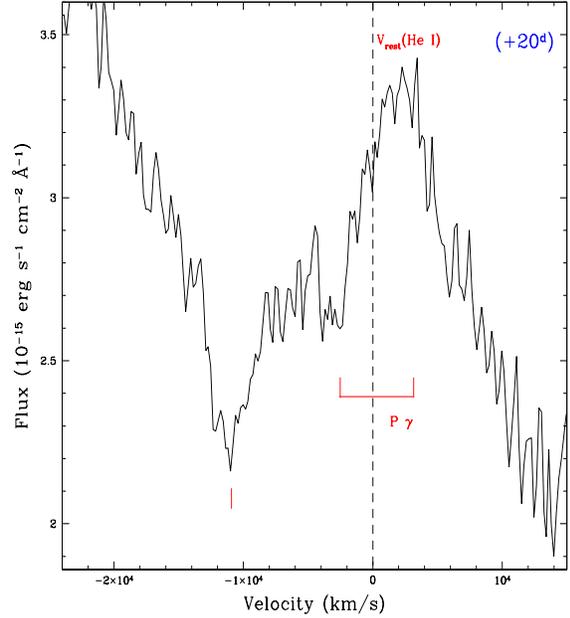,width=8cm,height=9cm}
\caption {He I 10830 \AA\ on day 20. The line is blended with P$\gamma$ which strongly affects the He I emission component, but does not 
affect the absorption component of 10830 \AA\ line whose absorption minimum is 
shown by a vertical tick at a velocity of $\approx 11000$ km s$^{-1}$. The expected positions of absorption and emission components of P$\gamma$ are indicated with the ``U'' shape segment. Note that P$\gamma$ and He I 10830 \AA\ velocities are different. The vertical short-dashed line corresponds to the He I 10830 \AA\ rest wavelength. }
\end{figure}
\subsection{Infrared spectra}
The infrared spectra are shown in Figure 9 where some line identifications are given. The first spectrum was obtained during the photospheric phase ($\sim+20~ \rm d$) while the second one corresponds to $\sim +444~\rm d$ after explosion during the nebular phase. The earlier spectrum is dominated by a strong continuum with clear evidence of the Paschen series of hydrogen, namely P$\beta$, P$\gamma$ and P$\delta$ displaying P-Cygni profiles as did the optical Balmer lines. The structure of the P$\gamma$ profile results from blending with He I 10830${\AA}$. The position of the peak emission and the absorption minimum of P$\gamma$ are reported in Figure 10, using a linear interpolation between the respective positions for P$\delta$ and P$\beta$. Note the difference in velocity.

The velocity of He I 10830 \AA\ absorption component is in fact similar to
that of H$\alpha$ estimated by interpolation between days 10 and 
41. The presence of He I 10830 \AA\ at this epoch may be a result of 
the non-thermal excitation due to the presence of small amount of 
$^{56}$Ni in the outer layers of the ejecta. Alternatively, excitation can result from the recombination of frozen ionization,
the mechanism suggested by Utrobin \& Chugai (2002). The second spectrum is dominated by emission lines as the absorption components of the P-Cygni profiles became weaker.\\ \hspace*{0.4cm} Combining these IR spectra together with optical spectra obtained at similar phases we have extended spectra at two quite different phases. Figure 11 displays the resulting spectra, after being corrected for redshift of the host galaxy as well as for reddening, together with three black-body fits to the photospheric phase spectrum corresponding to different temperatures, yielding a temperature of the order $5500$ K as the best fit. Both combinations required no arbitrary shifts between optical and IR sections. 
\subsection{H$\alpha$ evolution,``Bochum event" and $^{56}$Ni asymmetry}
The evolution of the H$\alpha$ profile from the early photospheric epoch 
through the developed nebular stage is in many respects similar to
that of SN~1987A: early photospheric epoch with normal P-Cygni
lines expected in the spherically-symmetric case, 
 development in the H$\alpha$ line of a fine structure resembling ``Bochum-event" in SN~1987A
at the late photospheric epoch, and the development of fine structure and 
red-shift in the emission maximum of H$\alpha$ at the nebular epoch (Fig.
12).

The first two spectra show an undisturbed P-Cygni profile of H$\alpha$ with a clear blueshift of the peak emission, in contrast with the standard picture of line formation in an expanding atmosphere that generally requires an unshifted emission maximum. Similar behaviour is also seen in spectra of other SNe II, namely SN IIP 1988A \shortcite{tura2}, SN IIL 1990K \shortcite{capp} and SN 1987A for which this behaviour has been a subject of theoretical studies and has been explained as being due to the reflection of photons by the photosphere (Chugai 1988; Jeffery $\&$ Branch 1990). There is a similar blueshift in the emission peak of H$\beta$, clearer for the first spectra. Figure 13(left) displays the amount of the blueshift seen in our spectra. 
\begin{figure}
\psfig{file=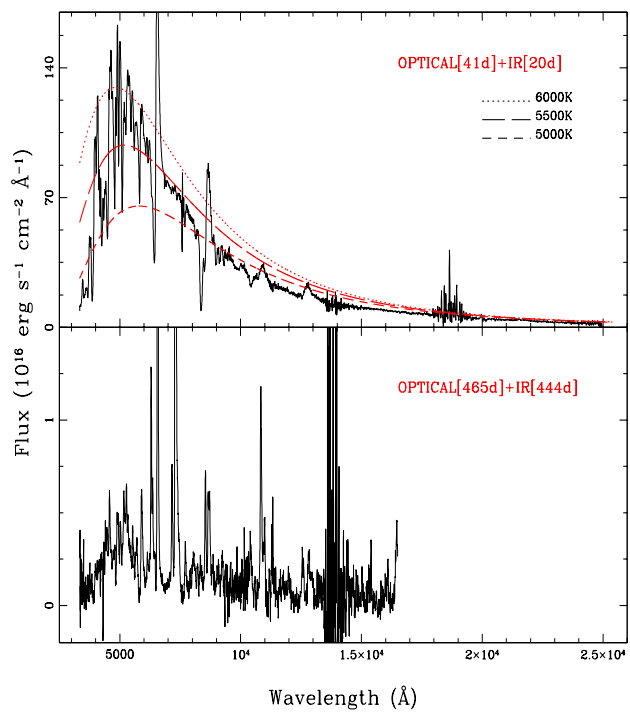,width=9cm,height=10cm}
\caption{ The resulted extended spectra from the best combination of infrared and optical spectra. Both spectra are corrected for redshif. The earliest OP($+41~\rm d$)+IR($\+20~ \rm d$) spectrum, shown on the top, is also corrected for reddening. Black-body curves at 3 different temperatures are plotted with the best fit corresponding to a temperature of the order $\sim5500$ K .}
\end{figure}
\begin{figure}
\centerline{\psfig{file=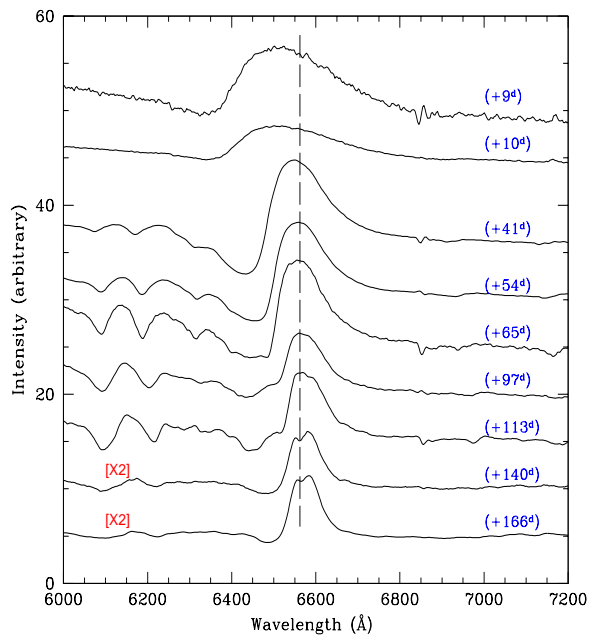,width=8cm,height=9cm}}
\caption{ The evolution of H$\alpha$ spectral region. The spectra are corrected for the host galaxy redshift and scaled arbitrary in intensity, some spectra are also multiplied by a shown factors for clarity. The rest wavelengths and days since explosion are also reported .}
\end{figure}
As the supernova evolves, the blueshift decreases, so that by $\sim65~\rm d$ it approaches zero, more clearly for H$\alpha$ than for H$\beta$ where line blending is a complication. The blueshifts for the Paschen hydrogen lines from our first infrared spectrum are also shown to be in good agreement with H$\alpha$, indicating similar properties of the photosphere in the optical as well as in the infrared. 

In Figure 13(right) we display the evolution in time of the H$\alpha$ velocity of SN 1999em together with that of SN 1987A. Velocities are derived from the minima of the P-Cygni absorption components, and in SN 1999em are lower than in SN 1987A at similar phases. The nature and the similarity of the evolution confirm the initial rapid decline in velocity corresponding to the rapid cooling period, followed by a slowly decreasing velocity period as expected during the plateau phase.
The expansion velocities at $\sim97~\rm d$ and $\sim113~\rm d$ and also to a lesser extent at $\sim65~ \rm d$ deviate from an otherwise smooth trend. This is due to the emergence of some structures in the absorption components that make the determination of the minimum absorption complicated and doubtful. The evidence of some structures in SN 1999em spectra has been noted first by Leonard et al. \shortcite{leo1}. 

Around $\sim65~\rm d$ we note some flattening of the absorption component with a hint of a weak peak near 6496 \AA\ in the transition between the minimum and maximum flux. This subsequently becomes more pronounced. A less pronounced feature develops on the red side of the emission component at $\sim97~\rm d$ near 6675 \AA. All these features show a tendency to smaller velocities (i.e. they move toward the rest wavelength) as time increases, reaching their maximum prominence around $+113~\rm d$. Similar but not so clear behaviour is evident for other lines in SN 1999em, in particular Na I D. Figure 14 illustrates the structure and the presence of these ``bumps'' on the lines at day 113. On the H$\alpha$ profile we mark three detectable structures: 1 blue and 2 red bumps. In the lower plot we show the Na I D profile which reveals clearly the presence of two red bumps at velocities very close to r$_{1}$ and r$_{2}$ seen in H$\alpha$. Moreover we note that the b$_{1}$ feature velocity is in agreement with the photospheric velocities as derived from weak metal lines (Fe II 5018  \AA\ and 5169 \AA) at corresponding time. For the  $+113~\rm d$ spectrum : $V_{\rm Fe~II}\sim 2100$ km s$^{-1}$ while the corresponding velocity from the peak of the b$_{1}$ feature is of the order $V_{\rm b1}\sim2400$ km s$^{-1}$. 

This complicated structure of the H$\alpha$ profile was also noted for SN 1987A (H$\alpha$ fine structure and Bochum event; Hanuschick 1988; Phillips et al. 1989; Sartori et al. 1990) and in SN 1988A (Turatto et al. 1993). It occurred earlier in SN 1987A and has been ascribed to clumping and mixing of radioactive material into the outer envelope with an axially symmetric geometry, two-sided jet, by  Lucy (1988). This possibility is illustrated by Figure 15(left) which shows the spectrum on day 97 with the model H$\alpha$ computed assuming 
a spherical distribution of optical depth and net emission source functions.
The residual shows excess in the blue and red, which conceivably reflects 
the overexcitation produced by $^{56}$Ni jets.
It is, of course, not a unique representation of the spherical model, but this one at least demonstrates this possibility. An alternative view on the fine structure in H$\alpha$ of SN~1987A 
suggests that the blue bump, on the contrary, reflects underexcitation
of hydrogen at some velocity in the atmosphere related to 
the deep recombination following the cooling wave (Chugai 1991a). 
Recently this conjecture was confirmed by 
the non-steady state calculations of the hydrogen recombination 
in the atmosphere of SN~1987A with the inclusion of reactions with 
H$^-$ and hydrogen molecules \shortcite{utro}.
This modelling also confirmed the dominant role of the ionization 
freeze-out effect in the population of the second hydrogen level and 
the H$\alpha$ formation at the photospheric epoch \shortcite{chug2b}.
According to the latest computations the non-monotonic behaviour of the excitation 
in the atmosphere of SN~1987A (and SNe~IIP in general) would be 
a natural outcome of the freeze-out effects, combined with the 
neutralization of H$^+$ in the ion-molecular reactions.
Of course, only direct similar modelling for SN~1999em will give a 
conclusive answer concerning the plausibility of this mechanism for 
the blue bump. 
Figure 15(right) demonstrates, how the Sobolev optical depth in H$\alpha$ 
would appear in order to reproduce the appropriate blue bump. 
Note, although the blue structure is fitted quite well, some 
excess in the red remains unaccounted for in the model and therefore indicates 
a real asymmetry of the hydrogen excitation. 
\begin{figure*}
\centerline{\psfig{file=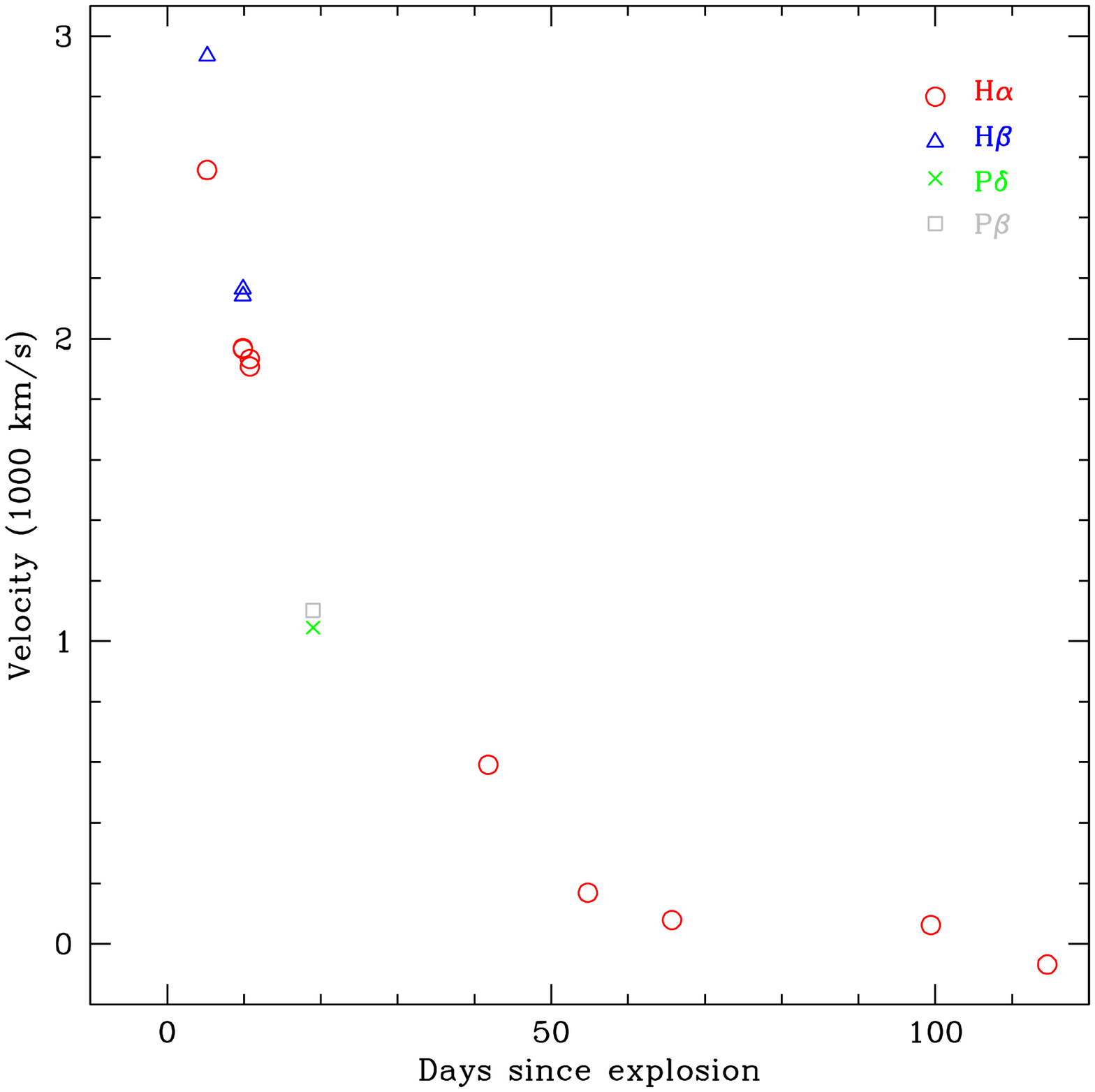,width=7.5cm,height=9cm}\psfig{file=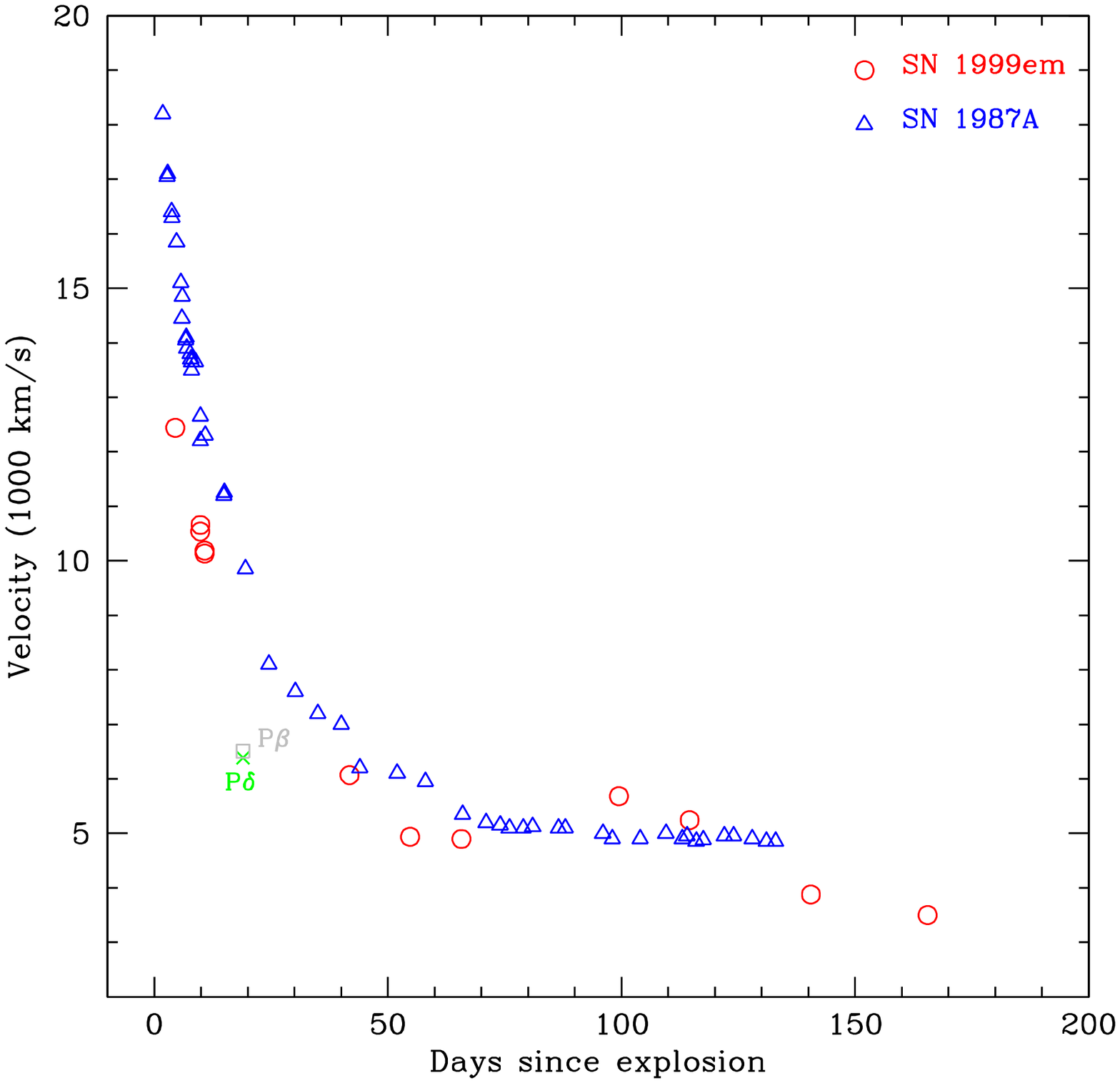,width=7.5cm,height=9cm}}
\caption{ Left panel: the amount of the blueshift seen in the emission peaks of H$\alpha$ and H$\beta$ during the first phases. Line blending is a complication for H$\beta$. Right panel: the evolution of the expansion velocities, derived from the absorption minima in H$\alpha$ profiles, for SN 1999em together with that of SN 1987A for comparison. For both panels the corresponding velocities of P$\beta$ and P$\delta$ from the $\sim+20~\rm d$ infrared spectrum are also shown.}
\end{figure*}
\begin{figure}
\centerline{\psfig{file=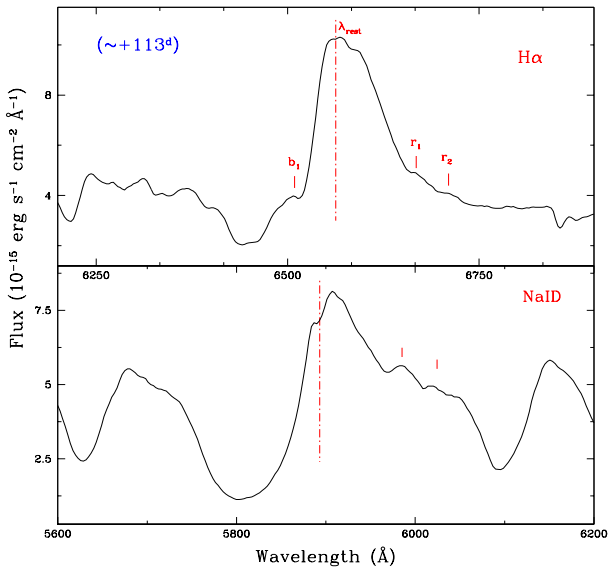,width=8.5cm,height=8.2cm}}
\caption{The H$\alpha$ and Na I D line profiles at $\sim+113~\rm d$ after explosion. The rest wavelengths are also shown. The short bars in Na I D profile correspond to the derived velocities from the features seen in the H$\alpha$ profile (see discussion in the text). }
\end{figure}

In fact soon after, on day 113 the red excess became even more 
  evident. It increases with time and on day 444 it is also detected in He I 10830 \AA. On days 140 and 166 the H$\alpha$ maximum shows significant 
deviation from a round-shaped form. The complicated structure 
with the dominant red peak apparently indicates the 
asymmetric distribution of the line-emitting gas in the 
central region $v<1500$ km s$^{-1}$.
The comparison of He I 10830 \AA\ with H$\alpha$ at the similar epoch (Figure 16) shows 
that the peak is redshifted by the same value as in H$\alpha$.
We believe that this red bump at $\approx+400$ km s$^{-1}$ is 
caused by the $^{56}$Ni asymmetry.
The He I line is narrower (in the red P$\gamma$ contributes).
The blue width at half maximum (BWHM) is 2000 km s$^{-1}$ for 
H$\alpha$ and only 1000 km s$^{-1}$ for He I 10830 \AA.
The He I 10830 \AA\ line traces non-thermal excitation more closely than H$\alpha$, since for hydrogen additional ionization from the second level is more important than for He I. The width and position  of the He I 10830 \AA\ line thus indicates that the bulk of $^{56}$Ni is distributed inside a sphere with velocity $v<1500$ km s$^{-1}$ and the $^{56}$Ni zone is shifted to the far hemisphere by roughly 400 km s$^{-1}$.
\section{Evidence for dust formation around day 500}
\begin{figure*}
\centerline{\psfig{file=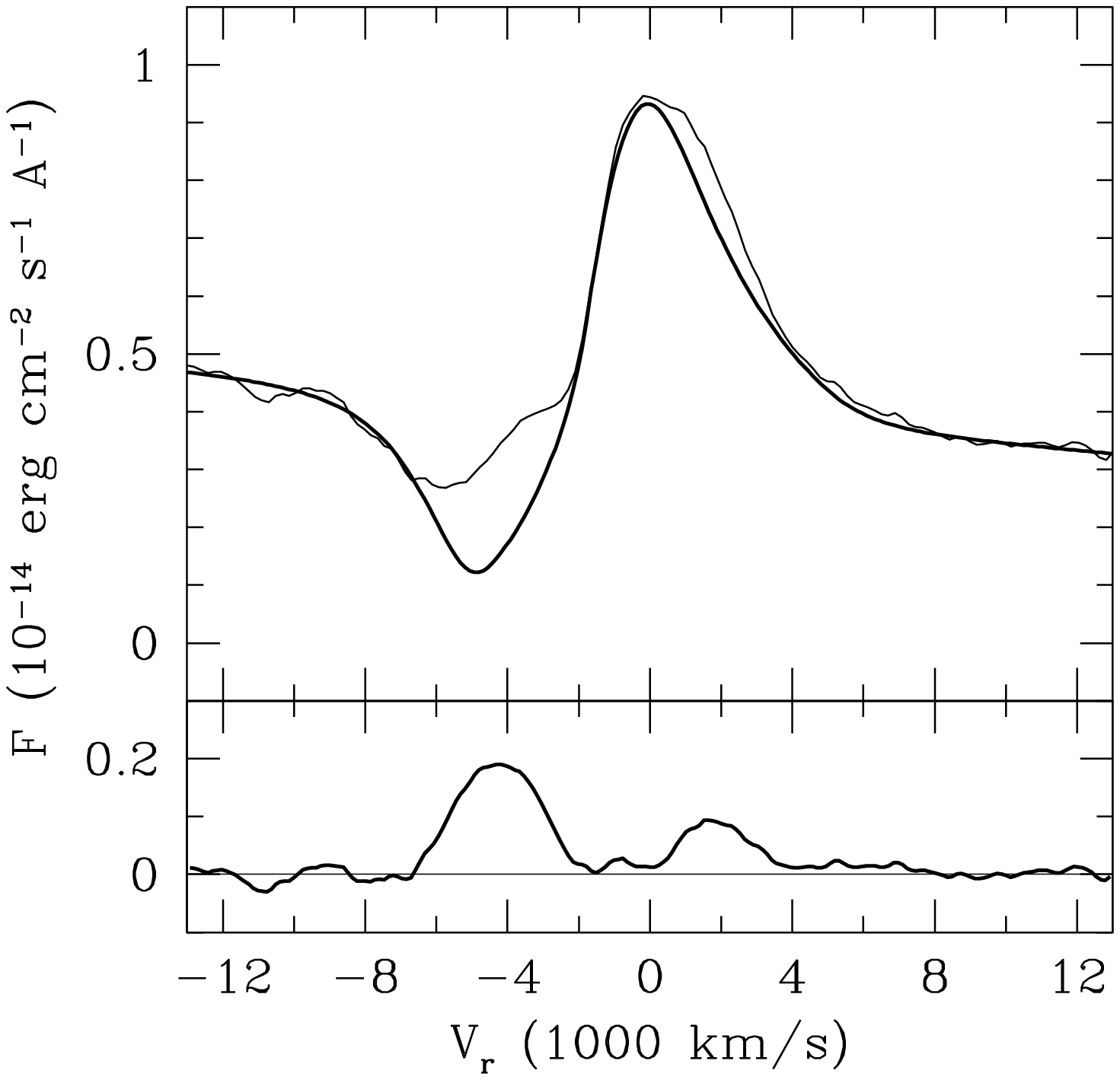,width=10cm,height=9.5cm}\hspace*{-1cm}\psfig{file=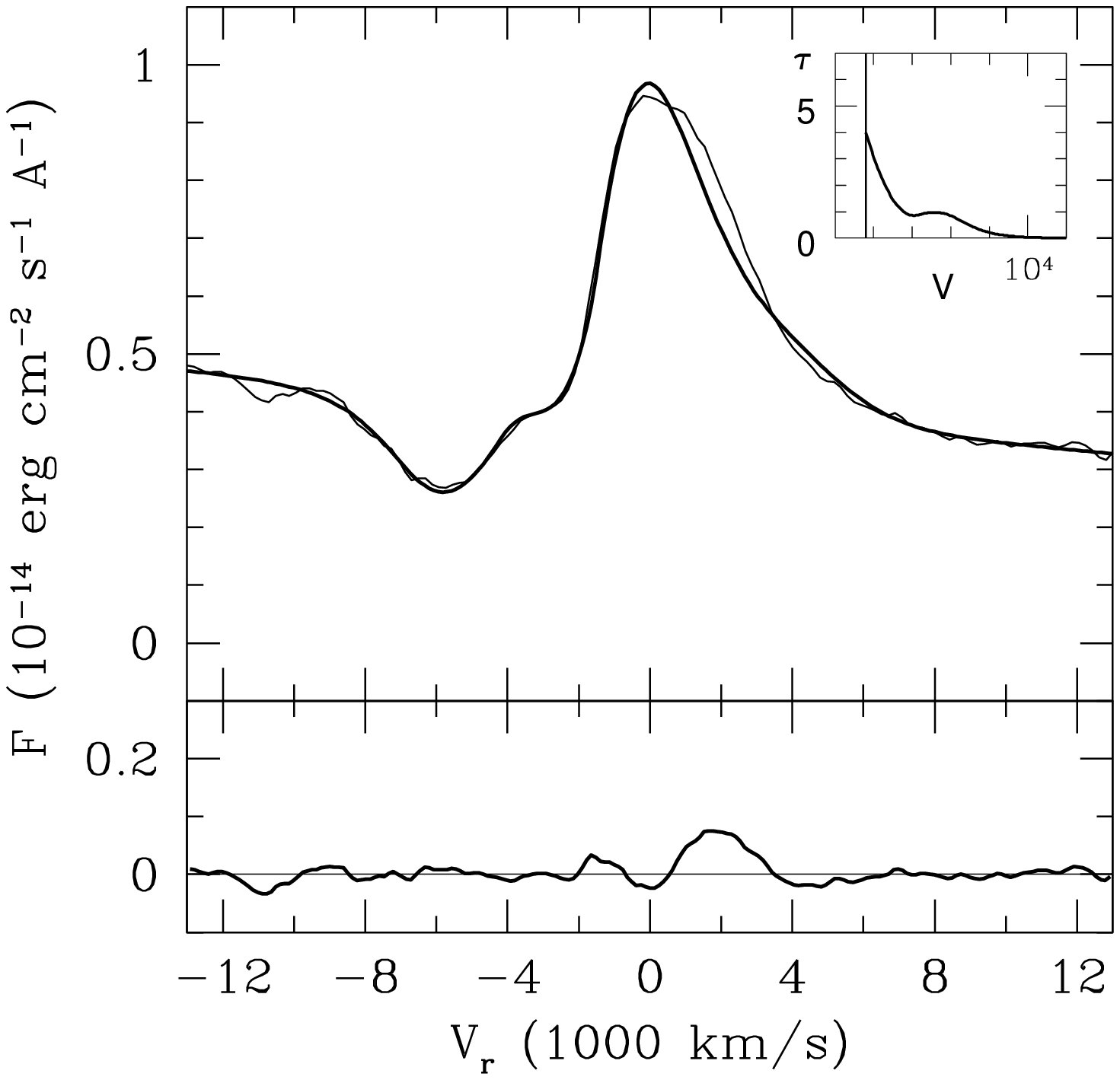,width=10cm,height=9.5cm}}
\vspace*{-1cm}
\caption{Left panel: the H$\alpha$ on day 97. The overplotted (thick line) is a
profile for a spherical model. The residual (bottom panel) shows
two peaks, which can be interpreted as evidence for the presence of 
bipolar excitation regions in the atmosphere. Right panel: the H$\alpha$ on day 97. The overplotted (thick line) 
is the profile for the spherically symmetric model with the 
non-monotonic behaviour of the optical depth shown in window 
(up right corner). The residual shows the excess in the red, 
which is indicative of overexcitation in the far hemisphere. }
\end{figure*}
The temporal evolution of the [O I] profile can be an effective indicator of dust formation in the form of an observable blueshift in the emission peaks of the components \shortcite{dan2}. It should be noted, that apart from the late time blueshift related to the dust formation, SN~1987A also showed an ``early
blueshift" in [O I] 6300 \AA\ between day 200 and 300 that 
disapeared at the later epoch. We believe, the blueshift 
in [O I] 6300 \AA\ observed in SN~1999em on day 312 and 391 
and which vanished on day 465 is of the same origin as the early 
blueshift observed in SN~1987A and also that reported for SN 1988H (Turatto et al. 1993). Namely, it has nothing to do with 
the dust and probably is the result of the superposition 
of the blend of Fe II lines (multiplet 74). This latter is hinted at by
the bump at 6250 \AA\ attached to the blue wing of [O I] 6300 \AA.

However between day 465 and 510 the [O I] 6300,6364 \AA\ line and H$\alpha$ 
demonstrate a pronounced evolution (Figure 17). 
The transformation in both profiles can be described as 
 a flattening of the line accompanied by skewing towards the blue. 
The quality of the 642 d spectrum is not 
 sufficient to confirm this effect, although 
 the spectrum does not contradict the conclusion 
 from the previous two spectra. 
The dramatic change of the line profile at the time scale 
of the order of 0.1 of the expansion time (or even faster)
 and a proximity of the age to the epoch of the dust formation 
 in SN~1987A around day $\sim 526$ \shortcite{lucy2} suggest that 
 dust formation is the likely cause of this profile transformation.
Note that the flattening of [O I] 6300 \AA\ line profile in fact was 
 not observed in SN~1987A at the dust formation epoch. Instead
  the blueing of oxygen line in SN~1987A preserved the  
 round-topped profile of the [O I] 6300 \AA\ line 
 which is consistent with the dust formation in the sphere with 
 the expansion velocity of $1800$ km s$^{-1}$ and dust optical depth
 of the order of unity (Lucy et al. 1989).

Nevertheless, the flat-topped profile is expected, if the optical depth 
 of the opaque core is very high.
To emphasize the point we present the toy model (Figure 18), which 
 suggests some smooth distribution of emissivity 
 in the [O I] doublet lines (with the line ratio of 0.4, similar to 
 SN~1987A at the same epoch) and the embedded 
 opaque core.
The model doublet in the absence of any absorption 
 roughly represents the situation on day 465. Switching on the dust 
 absorption in the dusty core with the boundary velocity of $800$ km s$^{-1}$
 produces the effect, which markedly depends on the optical depth.
For $\tau=2$ the line shape and blue shift recalls what was seen in SN~1987A at the dust formation epoch after day 530, where the estimated optical depth of the core is $\tau_{\rm d} \sim 1$ (Lucy et al. 1989). 
We see that the flat-topped [O I] doublet profile observed in SN~1999em on day 510 requires an optical depth of the dusty sphere 
 $\tau_{\rm d} \gg 10$, since in case of $\tau_{\rm d} = 10$ 
 the plateau has a noticeable inclination, which is absent in the case of 
 $\tau_{\rm d} = 50$.
This illustration indicates that the dust formed between days 465 and 510 and the optical depth of the dusty core was enormous, $\tau_{\rm d} \gg 10$.
The blue end of the plateau in the [O I] 6300 \AA\ line (Figure 17)
 indicates the velocity of the dusty sphere of about $v_{\rm d}\approx  800\pm 50$ km s$^{-1}$. 
\begin{figure}
\centerline{\psfig{file=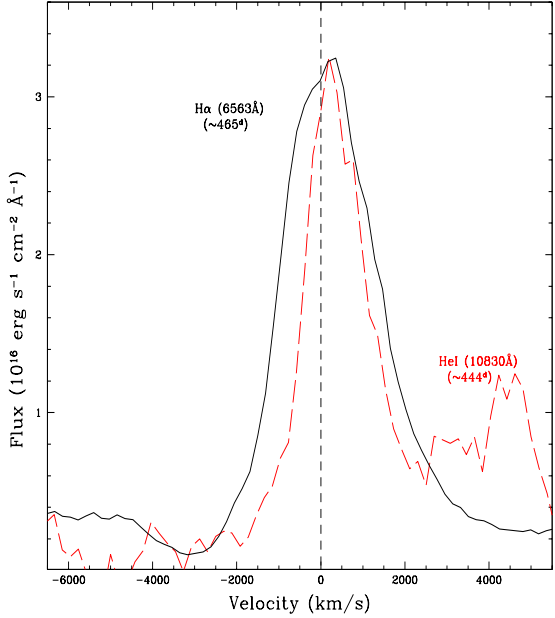,width=8cm,height=9cm}}
\caption{ H$\alpha$ and He I 10830 \AA\ lines at the nebular epoch. 
Both lines show apparent redshift indicating the asymmetry of $^{56}$Ni.
The He I line blue width at half maximum is factor two lower, possibly
 indicating the actual width of the $^{56}$Ni zone. Note that the He I 10830 \AA\ maximum is shifted arbitrary to reach H$\alpha$ maximum for comparison.}
\end{figure}

As in SN~1987A, for the H$\alpha$ line the effect of the 
 dust core is less apparent because of the more extended
 H$\alpha$-emitting zone and, primarily, because  of the redward asymmetry of this line related with the
   $^{56}$Ni asymmetry.
Nevertheless, the blueward skewing in the spectrum on day 510 
 is seen in H$\alpha$ also.

Even less apparent is the profile evolution of the [Fe II] 7155 \AA\ line. 
The absence of the skewing towards the blue in this line on day 510
 indicates that the unocculted fraction of asymmetric red component is strong enough to maintain the original redshift of this line seen on day 465. 
To illustrate this explanation we represent the profile 
of [Fe II] 7155 \AA\ as a combination of the spherically-symmetric
component and the component originating in the conic structure with 
the opening angle $2\alpha$ and the angle between line of sight and 
the cone axis $\theta$ (Figure 19, upper panel).
Adding to the model the dust core with $v_{\rm d}=800$ km s$^{-1}$ and 
 $\tau_{\rm d}=50$, we checked, if our choice for the asymmetrical 
 component is tolerated by the 510 day spectrum of [Fe II] 7155 \AA\ .
 The decomposition of the line profile into 
 symmetric and asymmetric component is, of course,  not unique,
 given the large noise in the 510 d line (the usual drawback for the
 inverse problem).
However, we found that for small inclination angle, $\theta<50^{\circ}$,
 the skewing towards the blue in the model is unacceptably large. 
One of the acceptable possibilities found by trial and error procedure 
 is shown in Figure 19(lower panel). In this case $\theta=67^{\circ}$ 
and $\alpha=40^{\circ}$. The line emissivity in the model cone is concentrated towards the cone axis and distributed along the radius with the broad maximum in the range 
 $1000 - 3000$ km s$^{-1}$. In this modelling  we followed a somewhat arbitrary but sensible requirement that the radial displacement of the maximum of the jet emissivity is minimal. The model with the dusty core obviously reproduces the major property of the profile on day 510, namely, the shift of the maximum towards the red. 
We thus conclude that the dust core implied by the [O I] doublet 
 transformation is not in conflict with the [Fe II] 7155 \AA\ line profile
 provided that the $^{56}$Ni jet, responsible for the red asymmetric component, has relatively large inclination angle $\theta\geq 60^{\circ}$. We do not insist on the uniqueness of the model of asymmetric distribution
of emissivity of [Fe II] 7155 \AA. Yet the one-sided asymmetry of emissivity 
 implied by the profile evolution is strikingly  
 consistent with the one-sided $^{56}$Ni distribution suggested
above by H$\alpha$ and He I 10830 \AA\ lines.
\begin{figure}
\vspace*{-1cm}
\centerline{\psfig{file=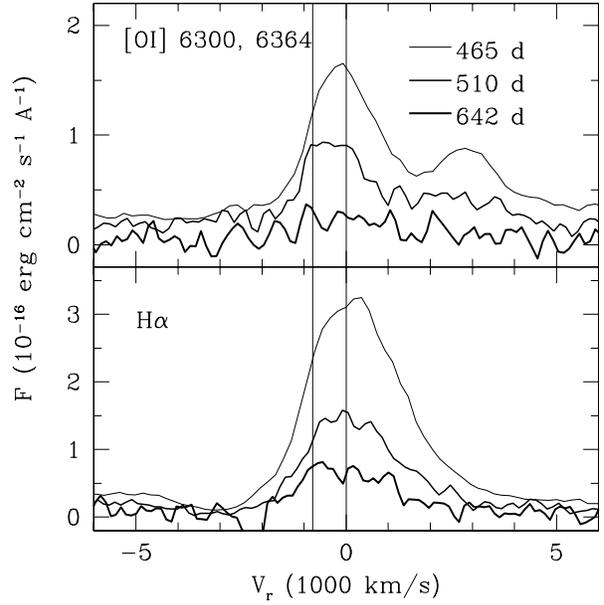,width=11cm,height=11.5cm}}
\vspace*{-1.5cm}
\caption{The oxygen doublet [O I] 6300, 6364 \AA\ and H$\alpha$ in 
the latest three epochs. The spectra on day 642 are multiplied 
by factor of six. The two vertical lines correspond to zero velocity 
and the approximate position of the blue edge of plateau in [O I] 6300 \AA\
line. The spectra are not corrected for reddening.} 
\end{figure}

The manifestations of the  dust formation in SN~1999em are 
 different from those in SN~1987A in at least three respects. 
First, while in SN~1999em the  [O I] 6300 \AA\ line revealed the 
 presence of the dust already on day 510, in the case of SN~1987A the
 blue shift in [O I] 6300 \AA\ emerged only after day 526 (Danziger et al. 1989; Phillips \&\ Williams 1991).
A somewhat earlier emergence of the dust in SN~1999em is probably 
 related to the lower amount of $^{56}$Ni, which in its turn 
 implies lower temperature at the given stage, so the condensation 
 temperature was attained in SN~1999em earlier.
Second, the velocity of the dust-forming zone is a factor two  
 lower compared to the dusty core in SN~1987A  ($\sim 1800$ km s$^{-1}$) 
 measured in the spectra around day 600 (Lucy et al. 1989). 
This difference possibly reflects both the smaller mass 
 (and therefore velocity) of the metal-rich core of ejecta
and smaller velocity of $^{56}$Ni bubble, which presumably pressurizes 
the metal-rich zone thus producing dense clumps in which dust forms.
The third is more clear cut: 
the optical depth of the dusty core in SN~1999em $\tau_{\rm d}\gg 10$ while 
 in SN~1987A $\tau_{\rm d} \leq 1$ \shortcite{lucy3}.
At first glance a factor two smaller radius of the dust core in SN~1999em compared to SN~1987A is an explanation for this difference. 
Assuming a comparable amount of dust and the given factor of two
 smaller radius one gets a factor of four higher column density.
However, this is still a small factor to explain at least   
 a factor of 20 difference in the optical depth.

The explanation of the high optical depth derives from the model of the clumpy dust zone in SN~1987A proposed by Lucy et al. (1991). To account for the wavelength 
 independent extinction the authors suggested that the dust is locked in 
  very opaque clouds, so the effective optical depth of the dusty 
  zone is actually the geometrical (occultation) optical depth produced 
  by the cloud ensemble  
  ($\tau_{\rm oc}= 3N/4\pi R_{\rm d}$, where $N$ is the number of clouds,
  $R_{\rm d}=v_{\rm d}t$ is the radius of dusty core).
Lucy et al. (1991) estimate  $\tau_{\rm oc}\approx 0.4$ for SN~1987A
 around day 625. 
To account for the high optical depth of the dusty core   
 in SN~1999em, we must admit that the occultation optical depth 
 is very high $\tau_{\rm oc}\gg10$, i.e. 
 the number of opaque cloudlets is substantially higher than 
 in SN~1987A.  
We estimate the amount of dust required to produce the optical depth 
 $\tau_{\rm d}=50$ on day 510 as $\sim 10^{-4}~M_{\odot}$, a quite
 moderate value. In fact, this number should be considered as a lower limit. An apparent blueward shift in the peak of Mg I] 4571 \AA\ is apparent between days 465 and
510. Since the shift is similar in velocity to that observed in [O I] 6300 \AA\ and
clearly results from a depletion of flux on the red side, with the blue side
remaining constant in velocity, it is consistent with our model for a very opaque core of dust with a large occultation optical depth.
\begin{figure}
\vspace*{-3cm}
\centerline{\psfig{file=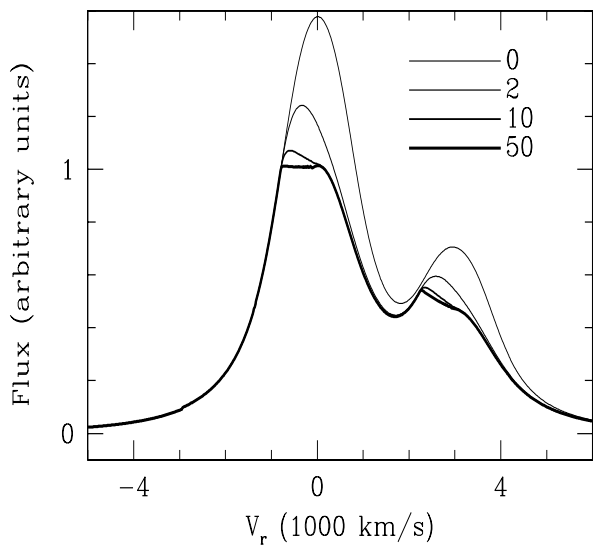,width=11cm,height=14cm}}
\vspace*{-3cm}
\caption{The effect of the opaque core on the [O I]  doublet. Lines of different thickness correspond to different optical depths of the core, starting with $\tau=0$ (thinnest line) through $\tau=2$, $\tau=10$, and $\tau=50$ with progressively growing line thickness. Note, for $\tau=10$ the plateau is still not settled and only for $\tau\gg10$ the profile is flat-topped.}
\end{figure}

The opaque dusty core must radiate as a black-body 
 with the temperature determined by the absorbed luminosity 
 and the radius. Modelling the H$\alpha$ profile transformation we 
 estimate that about 30\% of the luminosity is intercepted by the dusty 
 core on day 510. Given the bolometric luminosity at this epoch 
 $L\approx 2\times 10^{39}$ erg s$^{-1}$ the luminosity of the dusty 
 core must be then of $L_{\rm d}\approx 6\times 10^{38}$ erg s$^{-1}$.
The black-body temperature of the dusty core on day 510 was then 
 $\approx 510$ K with the maximum of the infrared spectrum at 6 $\mu$m.
We are not aware of any IR observations of SN~1999em 
 in M band at this epoch.
\begin{figure}
\vspace*{-1cm}
\centerline{\psfig{file=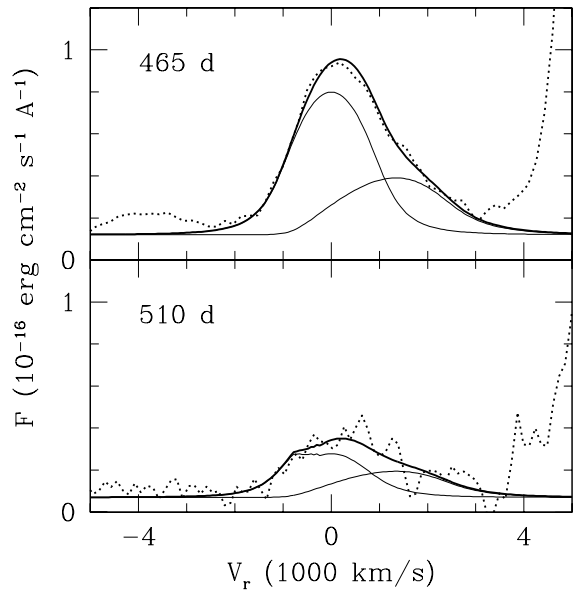,width=10.5cm,height=11cm}}
\vspace*{-1.5cm}
\caption{The effect of the opaque dusty core on the [Fe II] 7155 \AA\ line. Top panel shows the model line profile model ({\em thick line}) overplotted on the observed profile on day 465 ({\em dotted line}).
The model is a combination of the spherically-symmetric and asymmetric components ({\em thin lines}). The asymmetric component is modeled by the conic 
 overexcitation zone in the far hemisphere. The lower panel shows a similar plot but on day 510 and with added opaque core velocity $800$ km s$^{-1}$ and $\tau=50$ in the model. Note that the model flux on day 465 was multiplied by factor 0.5 to produce the adequate fit on day 510. }
\end{figure}
\begin{figure}
\centerline{\psfig{file=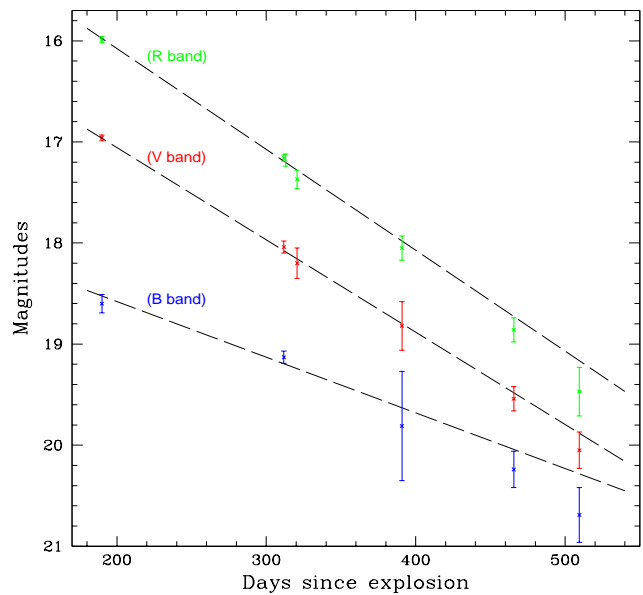,width=9cm,height=8.5cm}}
\caption{The late B,V and R light curves. The dashed lines represent the least squares fits of data weighted according
to error estimates and excluding the last point at day 510. A deviation is
seen in the last point of each colour, which supports the dust condensation
scenario after day 465 as inferred from spectroscopy.}
\end{figure}
Since vibration-rotation bands of CO have been detected in the IR spectra of
SN 1999em by Spyromilio et al. (2001), one is entitled to conclude, as was done
with SN 1987A, that the formation of molecules was a prelude to formation of
dust. Moreover we expect dust condensation to cause a deficit in the optical radiation with respect to the case with no dust. Figure 20 displays an expanded version of the B, V and R late time light curves together with the least squares fits (the last point is excluded in the fit). It is clear that all the light curves display a deficit decline for day 510 compared with the linear trend shown by fitting the previous data as represented by the dashed line. The average deficit is close to 30 percent. This gives further support for dust formation after day 465. Such a deficit was also noted by Lucy et al. (1991) at the time of dust formation in SN 1987A.

The fact that a fully opaque dusty core absorbs only a relatively 
small fraction ($\sim 30\%$) of the optical radiation is apparently 
related to the compactness of the dusty core and the large extent 
of the optical emitting zone. This is indicated by the
fact, that the velocity of the dusty sphere is notably lower than the FWHM
of H$\alpha$ on day 465 ($\approx 2700$ km s$^{-1}$).
A simple geometrical model to illustrate the situation is the dusty sphere with the velocity $v_{\rm d}$ embedded in the homogeneous extended emitting sphere
 ($v_{\rm e}$). It is easy to show then that the absorbed energy fraction  $\sim 0.3$ implies the ratio $v_{\rm d}/v_{\rm e}\sim 0.6$. Given $v_{\rm d}\approx 800$ km s$^{-1}$, this leads to the effective velocity of the emitting sphere  of $v_{\rm e}\approx 1300$ km s$^{-1}$. 

We may use a more physical model to directly constrain the extent of a $^{56}$Ni zone ($v_{\rm Ni}$) from the relative amount of absorbed light. Let the SN envelope be a homogeneous freely expanding sphere with the outer velocity $v_0$ determined by the ejecta mass $M$ and kinetic energy $E$. We assume that $^{56}$Ni is distributed homogeneously in the range $v<v_{\rm Ni}$. We may then compute the bolometric luminosity (without infrared dust emission) treating the gamma-ray transfer in the absorption approximation with the conventional effective absorption coefficient $k=0.03$ cm$^2$ g$^{-1}$. The positron deposition is assumed to be local. Simulations with and without dust using $v_{\rm Ni}$ as a tuning parameter permit us to find the extent of the $^{56}$Ni zone from a condition that the dusty core with $v_{\rm d}=800$ km s$^{-1}$ and $\tau_{\rm d}=50$ absorbs $30\%$ of the light generated locally with the deposition rate. In the particular case of the ejecta mass of $M=12~M_{\odot}$, and two values of kinetic energy of $E=10^{51}$ erg and $E=5\times10^{50}$ erg we find that the required radius of the $^{56}$Ni zone must be $v_{\rm Ni}=1100$ km s$^{-1}$ and 1300 km s$^{-1}$, respectively. The parameter variation probably would not change the results markedly, although an asymmetry could slightly affect the situation. The resulting extent of the $^{56}$Ni zone ($\sim 1100-1300$) km s$^{-1}$ is quite sensible and consistent with our previous finding that $^{56}$Ni lies inside the sphere of $\sim 1500$ km s$^{-1}$. Remarkably, the above simple geometrical analysis led to the extent of the emission region ($1300$ km s$^{-1}$) comparable to the extent of $^{56}$Ni zone in the dense (low energy) model. This reflects a simple truth that a dense model is close to the situation of the local deposition, the approximation implicitly assumed in the simple geometrical model. We therefore conclude that the small partial blackout of the supernova light following the formation of an extremely opaque dusty core of SN~1999em stems from the fact that the dusty core is compact compared to the extended distribution of $^{56}$Ni.

\begin{figure*}
\centerline{\psfig{file=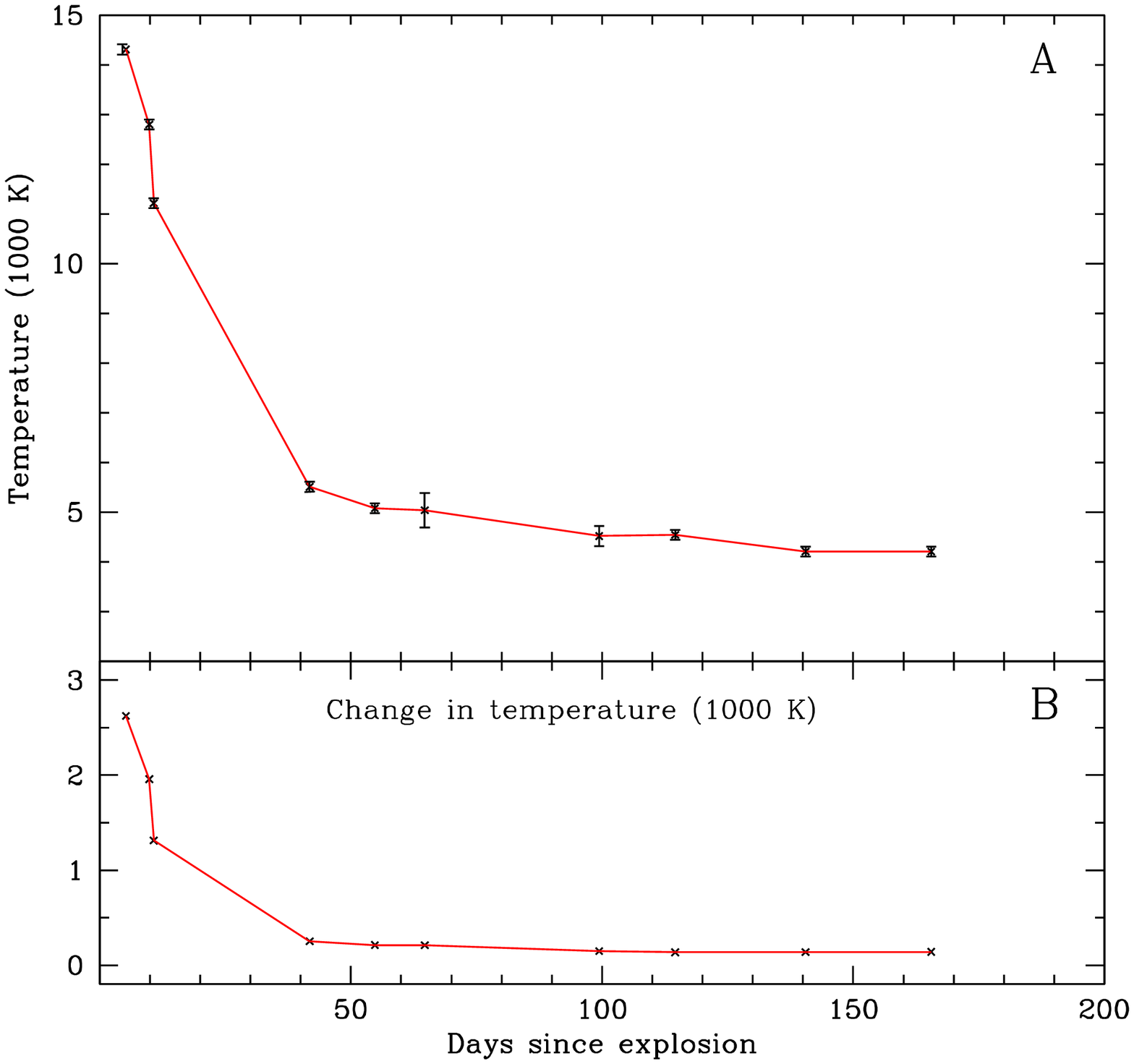,width=8cm,height=10cm}\psfig{file=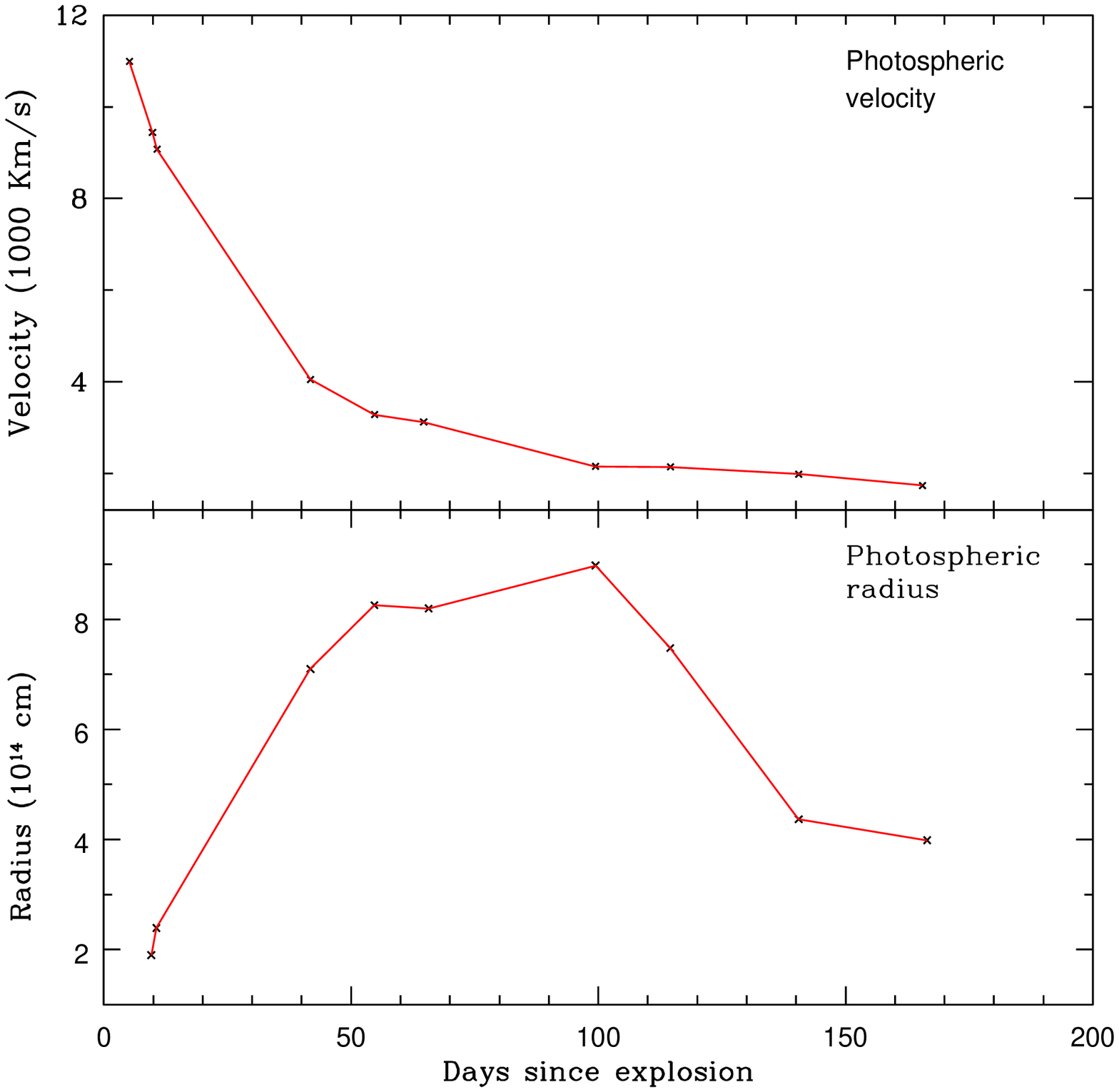,width=8cm,height=10cm}}
\caption{Left panel: the evolution in time of the derived temperatures from black-body fitting of the spectra (A), and the variation of the change in temperature from $\rm E(B-V)$=0.0 to $\rm E(B-V)$=0.1 (B). Right panel: the photospheric velocity and the photospheric radius evolution. This later is computed using the temperature and the bolometric luminosity (sections 2 $\&$ 5). }
\end{figure*}
\section{The photospheric temperature, explosion time and distance}
At the early photospheric epoch the photosphere of SNe~IIP is considered to approximate a black-body with a 
 sharp boundary \shortcite{bran}.
We made black-body fits to our spectra, after dereddening the fluxes in different bands with $\rm E(B-V)$=0.1. The interstellar extinction laws of Cardelli et al. \shortcite{card} were adopted. The derived temperatures are shown in Figure 20(upper panel), plotted as a function of time since explosion. 
We have used our Nov 3 spectrum to estimate the difference in temperature obtained by using a reddening of $\rm E(B-V)$=0.1 and the smaller reddening used by Baron et al. ($\rm E(B-V)$=0.05). That variation is then applied to the Baron et al. estimate for Oct 29 spectrum to give a revised higher temperature of T$\sim$14300 K.
In Figure 21(left lower panel), we plot the evolution of the difference in temperature between the case of no reddening correction ($\rm E(B-V)$=0.0) and the one with $\rm E(B-V)$=0.1. This behaviour shows the effect of the reddening correction on the derived temperatures from black-body fitting. At early phases, close to the explosion time and when the object is very hot, the colours on the Rayleigh-Jeans tail are not sensitive to temperature and therefore small changes in colour produce large changes in temperature and as time progresses this difference decreases. Figure 21(left upper panel), on the other hand, confirms the strong initial cooling of the photosphere as the supernova ages followed by a relatively constant temperature starting at T$\sim$5500 K and dropping to T$\sim$4200 K.\\
We apply the expanding photosphere method (EPM) to our data in order to estimate the time of explosion of SN 1999em and its distance. More detailed studies concerning the use of this method in the case of SN 1999em were made by Hamuy et al. (2001) and by Leonard et al. \shortcite{leo2}, who used broad band colours rather than spectrophotometry, and a larger data base.
The EPM, also called Baade-Wesselink method, is 
based essentially on the assumptions that the SN~IIP photosphere 
is spherical and emits as a dilute black-body, and that the 
expansion proceeds homologously ($v=r/t$). For small redshift 
($z\ll 1$) the photospheric angular radius is
\begin{equation}
\hspace{2truecm} \theta = \frac{R}{D} = \sqrt{\frac{f_{\rm \lambda} 10^{0.4 A_{\rm \lambda}}}{\xi_{\rm \lambda}^{2} \pi B_{\rm \lambda}(T)} } 
\end{equation}
where $R=vt$ is the photospheric radius determined by the expansion time 
$t$ and velocity at the photosphere $v$ (here we assume that the 
presupernova is a point),  $D$ is the distance to the SN, $f_{\rm \lambda}$ is the apparent flux density, $A_{\rm \lambda}$ is the total reddening coefficient, $B_{\rm \lambda}(T)$ is the Planck function evaluated at $T$, and $\xi_{\rm \lambda}$ is the dilution parameter to account for the fact that the SN does not radiate as a perfect black-body. 

We adopt $A_{\rm V}=0.31$, $f_{\rm \lambda}$ at $\lambda=5500$\AA\ (from observations), $B_{\rm \lambda}(T)$ using the temperatures of the black-body fitting of the spectra, and the parameter $\xi$ according to temperature and wavelength dependence given by 
to Hamuy et al. (2001).
The expansion velocity $v$ was measured from our spectra, using the absorption minima of weak lines such H$\beta$ and Fe II lines at 5018 \AA\ $\&$ 5169 \AA, formed in material close to the photosphere \shortcite{eastm}. The velocity evolution is shown in Figure 21(right). This leaves two unknowns, $t_{0}$ and $D$, to determine. Results are reported in Table 3. Figure 22 shows a least squares fitting of our data. We found $t_{0}\sim$ JD 2451476($\pm1$day) corresponding to 1999 October 24.5 and a distance of 7.83 $\pm$0.3 Mpc. These derived values agree well with results of the more detailed studies of the EPM for the case of SN 1999em (Hamuy et al. 2001 and Leonard et al. 2002). Our use of the same values of $\xi$ and the reddening as those used by Hamuy et al. links the two results. It is the use of different type of photometry and independantly determined velocities that strengthens the conclusions by both groups that the results are robust.
\begin{table*}
\begin{minipage}{140mm}
 \caption{The EPM results for SN 1999em.} 
\begin{tabular}{c c c c c c  }
\hline \hline
$Date$ (JD 2451000+) & $Flux$ ($\rm 10^{-14}ergs/s/cm^{2}/\AA$ in V)& $T_{\rm bb}~(\rm K)$ & $V$ ($\rm km/s$) & $\xi$ & $\theta~ (10^{13} $cm/Mpc$)$    \\ 
\hline
481.2 &1.101 & 14312 & 10990 & 0.45 & 6.034  \\
485.863 & 1.398 & 12802 & 9440 & 0.424 & 8.278 \\
486.767 & 1.385 & 11217 & 9072 & 0.42 & 9.685 \\
517.795 & 1.228 &  5510 & 4050 & 0.65 & 20.707 \\
530.755 & 1.184 & 5080 & 3280 & 0.75 & 21.578 \\
540.7 & 1.152 & 5040 & 3120 & 0.8 &20.361 \\
575.42 & 0.874 & 4520 & 2150 & 0.88 & 21.772 \\
\hline \hline 
\end{tabular} 
\end{minipage}\\
\hspace{-7truecm}\emph{\rm The values of $\xi$ are taken from Hamuy et al. (2001).}
\normalsize
\end{table*}
\begin{table*} 
\begin{minipage}{140mm}
\caption{Decline rates for the second plateau feature.}
\begin{tabular}{c c c c c }
\hline \hline
 Supernova & $B$ band & $V$ band & $R$ band & $I$ band \\
 & (mag)&(mag)&(mag)&(mag)\\
\hline
1997D &$\sim0.33$ in $221~\rm d$ &$\sim0.69$ in $89~\rm d$  & $\sim0.57$ in $79~\rm d$ &$\sim0.53$ in $51~\rm d$  \\
1991G &-----&$\sim0.3$ in $30~\rm d$  & $\sim0.2$ in $40~\rm d$ &$\sim0.15$ in $40~\rm d$ \\
1999em&$\sim0.22$ in $58~\rm d$ &$\sim0.5$ in $58~\rm d$  & $\sim0.45$ in $58~\rm d$ &$\sim0.54$ in $58~\rm d$  \\
\hline \hline
\end{tabular} 
\end{minipage}\\
\hspace{-5truecm}\emph{\rm The B-light curve of SN 1991G is not available in the work \\\hspace{-9.5truecm}of Blanton et al. (1995)}.\\
\normalsize
\end{table*}
\begin{figure}
\centerline{\psfig{file=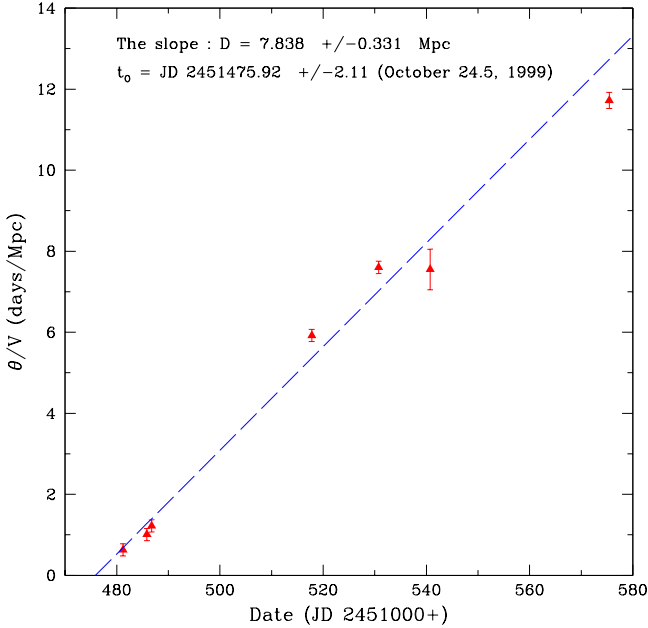,width=9truecm,height=9truecm}}
\caption{ The derived EPM results from our data. Also shown the least squares fit and the estimated explosion time and the distance to the SN. The discovery time corresponds to JD 2451480.94. }
\end{figure}
Thus the case of SN 1999em provides a check of the consistency of the EPM since, from observations, it is already known that the supernova exploded between Oct 20.45 and the discovery date Oct 29(IAUC 7294). In addition the derived distance is in excellent agreement with that derived from a study of the stellar component of NGC 1637 by Sohn and Davidge (1998).

We compute as well the photospheric radius using the luminosity and the temperature for a given time. The results are shown in Figure 21(right), together with the photospheric velocity evolution.
\section{Progenitor star properties }
The detailed observations of SN~1999em provide us with three important 
phenomenological parameters: the plateau duration ($t_{\rm p}$), 
absolute $V$ magnitude at the plateau ($M_V$) and the velocity at the
photosphere measured from weak absorption lines ($V_{\rm ph}$).
These three parameters permit us to recover 
the explosion energy ($E$), ejecta mass ($M$) and 
presupernova radius ($R$) using the light curve models as originally 
proposed by Litvinova \& Nadyozhin (1985).
In this paper useful expressions are given, which are used to 
estimate SN parameters. Similar, but somewhat different expressions 
are given by Popov (1993) based on the analytical model,
 which will also be used here. The input parameters of the model are the duration of the plateau phase ($t_{\rm p}$), a representative photospheric velocity at the beginning of the plateau phase ($V_{\rm ph}$) and its absolute $V$ magnitude ($M_{\rm V}$). From observations of SN~1999em we find $V_{\rm ph}\approx4050$ km
s$^{-1}$ using weak lines in the spectrum on day 41, $M_V\approx -15.76$ 
at the same epoch and the duration of the plateau $t_{\rm p}\approx 80$
days. According to the relations from Popov (1993) we derive the following
parameters for SN~1999em: ejecta mass $M\approx 11~M_{\odot}$, explosion
energy $E\approx 1.1\times10^{51}$ erg, presupernova radius $R\approx
120~R_{\odot}$. Applying the expressions derived by Litvinova \& Nadyozhin (1985) using a grid of hydrodynamic models for type II SNe, we find: ejecta mass $M\approx 9.8~M_{\odot}$, explosion
energy $E\approx 0.5\times10^{51}$ erg, presupernova radius $R\approx
180~R_{\odot}$. 
Both estimates are then consistent with a claim of a presupernova radius of
$\sim 120-150~R_{\odot}$, an ejecta mass of $10-11~M_{\odot}$ and 
an explosion energy $\sim(0.5-1)\times10^{51}$ erg. With approximately
$1.5~M_{\odot}$ enclosed in the neutron star and $0.5-1~M_{\odot}$ 
lost by the wind, we thus obtain the $12-14~M_{\odot}$ as a possible range 
for the main sequence mass of the progenitor. This is consistent with the 
values based upon oxygen mass estimate (section 3.1), and as well with the upper limit on the main-sequence mass using pre-supernova field images (Smartt et al. 2001).

The parameters derived using both sets of analytical expressions and the radius in particular ($\sim 120-150$ $ R_{\odot}$) suggest that the progenitor was a $G_{0}-G_{5}$ supergiant \shortcite{cox}. If it were at the detection limit of $m_{V}=23.2$ suggested by Smartt et al. \shortcite{smar} it would have had $M_{V}=-6.58$ and $M_{bol}=-6.42$ and an effective temperature of the order 5570 K. Using these parameters and the theoretical tracks of Maeder \& Meynet \shortcite{Maed} we conclude that the most probable mass would be 12$\pm$1 $M_{\odot}$. This conclusion is not significantly affected by uncertainties in rotational velocities and metallicity. These estimates provide one with the main parameters of the SN progenitor as a starting point for more detailed computations.

An interesting feature of the bolometric evolution of SN 1999em is the short duration of the peak before the fall to the plateau phase. In fact it has been shown that this early peak is very sensitive to the presupernova mass-loss history. Indeed Falk \& Arnett \shortcite{falk} have demonstrated that a rapid rise and short duration peak ($\sim$ 1-2 days) is consistent with the absence of an extended dense circumstellar shell and hence a low mass loss rate immediately prior to the explosion. This is also in agreement with $\it Chandra$ x-ray and radio observations of SN 1999em (Pooley et al. 2001).
\begin{table*} 
\begin{minipage}{140mm}
\caption{Parameters of some selected SNe.} 
\begin{tabular}{c c c c c c c}
\hline \hline
Supernova & $t_{\rm p}$& $E$ & $R$  & $M(^{56}\mbox{Ni})$ &$M_{\rm ms}$ & Source\\
   &(days) &($10^{51}$ergs) &($R_{\odot}$) &($M_{\odot}$) &($M_{\odot}$)\\

\hline \\

1999em & $\sim$ 80 & $\sim$ $0.5-1$& 120 - 150 & $\sim$ 0.022 & 12 - 14& A \\
1987A &$\sim$ 40 &$\sim$ 1.3 &$\sim$ 40  & 0.075&$\sim$ 20 &B\\
1997D & $\sim$ 50 &0.1& 85 & 0.002 & 8 - 12&C  \\
1993J & $---$ &1.6&$---$& 0.078 & 12 - 16 &D \\
1969L & $\sim$ 90 & $\sim$ 1.7 & $\sim$ 220 & $\sim$ 0.07 &$\sim$ 20&E \\
\hline \hline
\end{tabular}
\end{minipage}\\
\hspace{-4truecm}\emph{\rm \hspace*{3cm} A: The present work \hspace*{3.5cm} D: Utrobin 1996; Chugai \&\ Utrobin 2000\\\hspace*{-1.4cm}B: Danziger et al. 1988; Woosley et al. 1989 \hspace*{0.4cm} E: Arnett 1996; Sollerman et al. 1998 \\ \hspace*{-3cm} C: Chugai \&\ Utrobin 2000; (see Turatto et al. 1998 for alternative model)}\\
\normalsize
\end{table*}
\section{Summary and conclusions }
We have presented  photometric and spectroscopic data for SN 1999em, from $\sim9~\rm d$ until $\sim642~\rm d$ after the explosion. The shape of the light curve ($t_{\rm p}\sim$ 80 days) as well as spectral features show that it is a type IIP supernova. The problem of reddening has been discussed and we conclude that an optimum choice is close to that determined by Hamuy et al. (2001). The analysis of late phase photometry, up to $\sim510~\rm d$, shows that the exponential tail decay rate is close to the one of the radioactive decay $^{56}$Co to $^{56}$Fe, indicating that this is the main source of energy powering the light curve.

A photometric comparison of SN 1999em with SN 1987A, especially in the later phases, provides constraints on the radioactive $^{56}$Ni mass. We have constructed the ``UBVRI'' bolometric light curve, and comparing it with that of SN 1987A we obtain an estimate of the amount of  $^{56}$Ni produced by the explosion of 0.02 $M_{\odot}$, a smaller value than that derived for typical type IIP SNe such as SN 1969L and SN 1988A which have $M(\rm ^{56}Ni)\sim 0.07~\it M_{\odot}$.

We have noticed some flattening in the light curves, just after the steep decline from the plateau phase, and clearer for the blue bands. This behaviour is also seen in the light curves of two other objects, namely the peculiar SN IIP 1997D and the SN IIP 1991G. All three events have some common features, being type IIP and all having a lower ejected $^{56}$Ni mass: a similar amount for both SN 1999em and SN 1991G,($\sim$ 0.02 $M_{\odot}$) while SN 1997D ejected an even lower mass ($\sim$ 0.002 $M_{\odot}$). In addition, the duration of this $``$second plateau'' on the tail seems greater for SN 1997D than for SNe 1991G and 1999em. The decline rates and the observed duration of the flattening period are reported in Table 4. These measurements suggest that the $``$second plateau'' feature is a common feature for this low $^{56}$Ni mass type IIP supernovae, and that its duration is correlated with the amount of ejected $^{56}$Ni. Further support for this possible correlation comes from the case of SN IIP 1999eu (Pastorello et al. in preparation) which shows a very clear second plateau feature of duration $\sim 200~\rm d$ (clear in V and B bands). Moreover it seems that this SN has a very low ejected $^{56}$Ni mass similar to SN 1997D. Note that the prototype type IIP supernovae SN 1969L and 1988A do not show clear evidence of this behaviour. On the other hand SN 1969L and SN 1988A are known to produce an amount of $^{56}$Ni similar to SN 1987A $\sim$ 0.07 $M_{\odot}$. Improved statistical samples and better sampled light curves are required in order to confirm or rule out this behaviour. Radiation diffusion effects, still important at the beginning of the radioactive tail, are a possible cause of this behaviour. 

SN 1999em provided a test of the validity of the EPM since we have observational constraints on the explosion time and distance. In fact analysing the spectrophotometry, we derive an explosion time consistent with these constraints and in good agreement with what was found from more detailed broad band photometric studies of SN 1999em (Hamuy et al. 2001; Leonard et al. 2002).

We analysed the phenomenon of the fine structure of H$\alpha$ at the photospheric
 epoch, which was reminiscent of the ``Bochum event" in SN~1987A. Two possible 
 explanations, a
 bi-polar jet proposed by Lucy (1988) for SN~1987A and underexcitation of
 hydrogen combined with $^{56}$Ni asymmetry, are discussed. This analysis 
 does not permit one to discriminate between those models. Yet the one-sided 
  $^{56}$Ni ejection seems to find support in the small red shift of the He I 10830 \AA\ profile and the larger red shift of the H$\alpha$ line profile at the nebular epoch. We note that the He I 10830 \AA\ line should be more sensitive to non-thermal excitation resulting from $\gamma$-ray deposition. These lines 
  indicate that the $^{56}$Ni 
  distribution could be imagined to be a filled sphere with a velocity of 
  $\sim 1500$ km s$^{-1}$ shifted to the far hemisphere by 400 km s$^{-1}$.
 A somewhat surprising coincidence is that in SN~1987A the $^{56}$Ni distribution also 
 shows one-sidedness with a shift to the far hemisphere.

Analysing the [O I] 6300,6364 \AA\ line profile evolution we found a 
rapid change between days 465 and 510, which we interpret as 
an effect of the dust formation during this interval. Other support for dust formation comes from the deficit seen in optical radiation measured by late time photometry. It is the second SN~IIP (after SN~1987A) where convincing evidence of 
dust formation exists. In SN~1987A and SN~1999em we detected the 
rapid ($\Delta t/t\sim 0.1$) transformation of profiles during
the nebular epoch. The dust phenomenon in SN~1999em has some distinctive characteristics compared to 
SN~1987A. The dust condensation happened earlier (between days 465 and 510)
than in SN~1987A (after day 526), which is probably explained, by the 
lower $^{56}$Ni mass and, accordingly, lower temperature. The 
dust resides in the core with a velocity of $\approx 800$ km s$^{-1}$,
much lower than in SN~1987A. This greater confinement of dust in SN 1999em possibly results from the lower velocity creating a more confined metal-rich region where the condition for dust formation prevails. 
Another remarkable difference is the very large optical depth of the dusty zone ($\tau\gg10$) compared to SN~1987A ($\tau \sim 0.5$, Lucy et al. 1991).
We interpret this difference as an indication that in SN~1999em the dust
is distributed more homogeneously (or the number of the opaque 
dusty clumps is notably greater) than in SN~1987A. 

These facts show clearly the importance of studying more samples of SNe IIP (SNe II in general), because of the diversity they provide in manifesting the same event (i.e. dust condensation) and thus the opportunity of understanding the physics behind such events.

We used relations by Litvinova \& Nadyozhin (1985) and Popov (1993) 
to find SN parameters from observational characteristics. The estimated mass of the progenitor $M_{\rm ms}\approx 12-14~M_{\odot}$ 
and presupernova radius ($120-150~R_{\odot}$) are just consistent with 
the failure to detect a progenitor star with imaging of the pre-SN field. Moreover, our derived progenitor mass agrees with our finding that  
the oxygen mass in SN~1999em is about four times smaller compared to 
SN~1987A.

There is growing observational evidence of a significant decrease in the mass of Fe produced as the progenitor mass for type II SNe decreases when we include derivations for other objects. It is therefore encouraging to note that this behaviour is what is required to explain abundance patterns such as [O/Fe] and [Mg/Fe] in metal-poor halo stars modeled by Argast et al. \shortcite{argast}.

The correlation of progenitor mass of SN~IIP and core collapse SN in general with other parameters is vital for testing
  explosion models and the theory of stellar evolution. In Table 5, we present parameters of SN~1999em 
  along with other SNe~II.
 Note that the parameters reported in the table, except for SN 1999em, are not obtained directly by using the analytical models of Litvinova $\&$ Nadozhin (1985) and Popov (1993). They are however the most reliably determined through modelling of observations.

Table 5 demonstrates that for type II SNe there begins to emerge a monotonic
relation between progenitor mass on the one hand and both explosion
energy and $^{56}$Ni mass on the other. Although the uncertainties in these
parameters remain, of necessity, large, it seems that in the progenitor
mass range $10 - 13$ $M_{\odot}$ there is also a steep decrease in both energy and $^{56}$Ni mass. This suggests that these latter 2 parameters are not
independent but physically linked. Results for other core collapse
objects (type Ib,c) with higher progenitor masses tend to support this
correlation even if the current scatter is unavoidably large. Future
observations supported by modelling will surely elucidate this
conclusion. It is clear from our analysis that high S/N spectra of SNe at late phases are an invaluable tool for understanding type II supernovae.

The fact that SN 1999em probably resulted from the explosion of a $G_{0}-G_{5}$ supergiant also indicates, as did SN 1987A, that the evolutionary stage at which massive stars can explode is not yet well delineated.

\section*{Acknowledgments}
We thank E.Baron for providing the early spectrum and S.J.Smartt for providing us the 28 October 1999 image .\\
A.Elmhamdi is grateful to J.C.Miller at $SISSA/ISAS$ for useful discussion and comments. N.Chugai thanks INAF-OAT and Cofin 2000 for hospitality and support. We are also grateful to the VSNET observers for the online data.
We thank the referee, L.B.Lucy, for helpful comments.

\end{document}